\documentclass[submission,copyright,creativecommons]{eptcs}
\providecommand{\event}{MARS 2018} % Name of the event you are submitting to
\usepackage{breakurl}             % Not needed if you use pdflatex only.
\usepackage{underscore}           % Only needed if you use pdflatex.

\usepackage{graphics}
\usepackage{graphics}
\usepackage{amsmath}
\usepackage{makeidx}  % allows for indexgeneration
\usepackage{multirow}
\usepackage{graphicx,wrapfig,subfig}
\usepackage{array}
\usepackage{url}

\usepackage{makeidx}  % allows for indexgeneration
\usepackage{alltt}
\usepackage{graphicx,wrapfig,subfig}
\usepackage{pgfplots}
\pgfplotsset{/pgf/number format/use comma,compat=newest}
\usepackage{tgcursor}
\usepackage{xcolor}  %for recent updates marked in blue {\color{blue} xxx}
\usepackage{soul}  %for recent removals marked strikethrough  \st{xxx}

\title{Ten Diverse Formal Models \\for a CBTC Automatic Train Supervision System}
\author{Franco Mazzanti
\institute{CNR-ISTI\\ Pisa, Italy}
\email{franco.mazzanti@isti.cnr.it}
\and
Alessio Ferrari
\institute{CNR-ISTI\\ Pisa, Italy}
\email{alessio.ferrari@isti.cnr.it}
}
\def\titlerunning{Ten Diverse FMs for an ATS}
\def\authorrunning{F. Mazzanti \& A. Ferrari}
\begin{document}
\maketitle

\begin{abstract}
Communications-based Train Control (CBTC) systems are metro signalling platforms,  
which coordinate and protect the movements of trains within the tracks of a station, and between different stations. 
In CBTC platforms, a prominent role is played by the Automatic Train Supervision (ATS) system, which automatically 
dispatches and routes trains within the metro network. Among the various functions, an ATS needs to avoid deadlock situations, i.e., cases 
in which a group of trains block each other. In the context of a technology transfer study, 
we designed an algorithm for deadlock avoidance in train scheduling. 
In this paper, we present  a case study in which the algorithm has been applied. The case study has been encoded using ten different formal verification environments, namely UMC, SPIN, NuSMV/nuXmv, mCRL2, CPN Tools, FDR4, CADP, TLA+, UPPAAL and ProB. 
Based on our experience, we observe commonalities and differences among the modelling languages considered, 
and we highlight the impact of the specific characteristics of each language on the presented models.
\end{abstract}

\section{Introduction}
\label{sec:intro}
Communications-based Train Control (CBTC) systems are the \textit{de-facto} standard for metro signalling and control, 
including several interacting wayside and onboard components that ensure safety and availability of trains within the metro network. 
In the context of a technology transfer project named TRACE-IT, the authors of the current paper, together with representatives of a large railway company, designed 
one of the main components of a CBTC system prototype, namely the Automatic Train Supervision (ATS) system~\cite{ferrari2014commercial}. 
This is a wayside system that dispatches and monitor trains along the metro network, according to a set of predefined missions. 
The ATS includes a scheduling kernel, which shall ensure that, regardless of train delays, no deadlock situation occurs, i.e., the missions
are designed in such a way that it never happens that two or more trains block each other from completing their missions. 
In the context of the project, we applied formal methods to design and verify a scheduling algorithm that addresses the deadlock avoidance 
problem~\cite{mazzanti2014deadlock}. 
The application of the algorithm to the TRACE-IT case study was initially modelled and verified by means of the UMC tool~\cite{ter2011state,kandi,umcsite}. 
Then, the design of the case study was replicated with other six different formal frameworks -- i.e., SPIN~\cite{spin,spinsite}, NuSMV/nuXmv~\cite{nuxmv,smvsite}, mCRL2~\cite{mcrl2,mcrl2site}, CPN Tools~\cite{cpn,cpnsite}, FDR4~\cite{fdr3,fdrsite} and CADP~\cite{cadp,cadpsite} --  to explore the potential of  formal methods diversity~\cite{mazzanti2018towards}. 
This is the usage of different formal tools to validate the same design, to increase the confidence on the verification results~\cite{kuismin}. 
In the current paper, we present the models discussed in~\cite{mazzanti2018towards}, 
focusing on the differences between the modelling languages, rather than on formal verification diversity. Furthermore, we provide three additional models, using TLA+~\cite{tla,tlasite}, ProB~\cite{eventb,probsite} and UPPAAL~\cite{uppaalsmc,uppaalsite}. Within the context of this paper, our goal is to provide some feedback on the differences and traps that should be tackled when changing the reference frameworks, and the commonalities that would allow a simple translation from one framework to another. 
The models are made available in Appendix A and in attachment to this paper.

The remainder of the paper is structured as follows. In Sect.~\ref{sec:algorithm} we provide an overview of the modelled algorithm. 
In Sect.~\ref{sec:comparison} we present the different models, discussing commonalities and differences with a focus on syntactic and semantics discrepancies. 
Sect.~\ref{sec:conclusion} concludes the paper. In Appendix A, we report the different models presented. 

\section{A Deadlock Avoidance Algorithm for ATS}
\label{sec:algorithm}
\begin{figure*}[!htp]\vspace{-3 mm}
\centering
\includegraphics[width=1.00\textwidth]{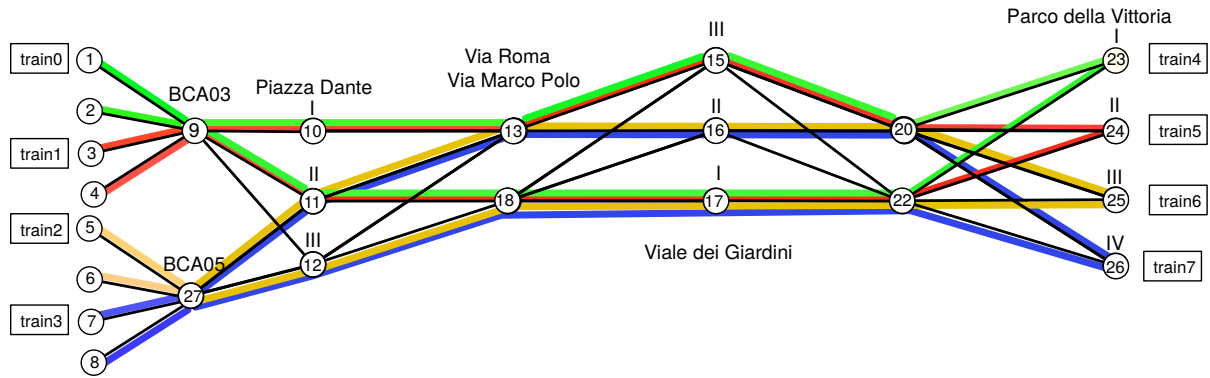}
\caption{A fragment of the yard layout and the 8 missions of the trains}
\label{fig:layout}
\vspace{-5 mm}
\end{figure*}

This section describes basic elements of the modelled algorithm, which was defined in our previous works~\cite{nfm14,mazzanti2014deadlock}.
Fig.~\ref{fig:layout} shows the structure of the railway layout considered in this study. Nodes in the yard correspond to itinerary endpoints, and the connecting lines correspond to the entry/exit itineraries to/from those endpoints. Eight trains are placed in the layout. Each train has its own mission to execute, defined as a sequence of itinerary endpoints.
For example, the mission of \texttt{train0}, which traverses the layout from left to right along top side of the yard, is defined by the mission vector: $T_0=[1,9,10,$ $13,15,20,23]$ (the numbers in the vector refer to the sequence of traversed endpoints in the diagram of Fig.~\ref{fig:layout}).
%The mission of train1, which also t verses the layout from left to right, is defined by the vector: T1=[3,9,10,13,15,20,24]. 
The mission of \texttt{train7}, which instead traverses the layout from right to left, is defined by the vector: $T_7=[26,22,17,18,12,27, 8]$.
The progress status of each train is represented by the index, pointing to a position in the mission vector, which allows the identification of the endpoint in which the train is at a certain moment.
We will have 8 variables $P_0, \ldots, P_7$, one for each train, which store the current index for the train. For example, at the beginning, we have
$P_0 = 0, \ldots, P_7 = 0$, since all the trains occupy the initial endpoints of their missions -- at index $0$ in the vector.
% The status of progresses for all the 8 trains can be represented by a vector P of 8 positions, where P[i] denoted the current progress of train i.
% The initial value for such vector is P=[0,0,0,0,0,0,0,0], i.e. all trains are in the first endpoint of their mission. 
%E.g., train0 is in endpoint 1 (T0[P0]) and train1 is in endpoint 3 (T1([P1]). 

\begin{figure*}[!htp]\vspace{-3 mm}
\centering
\includegraphics[width=1.00\textwidth]{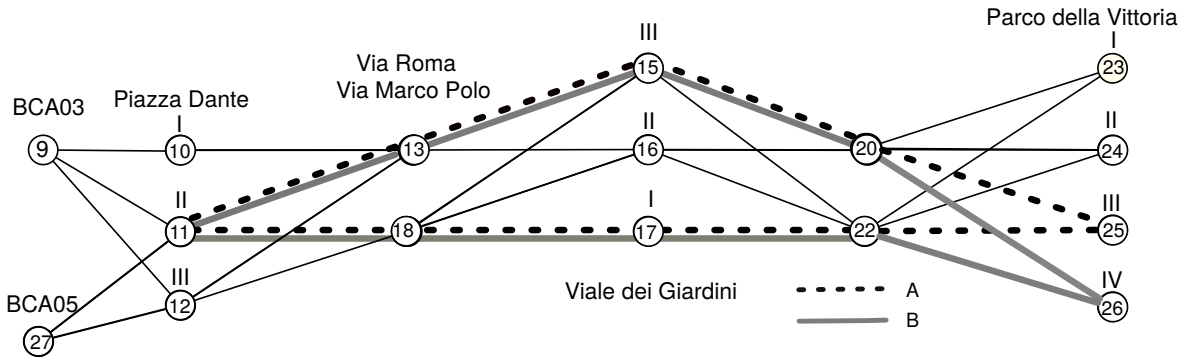}
\caption{\label{fig:criticalregion} The critical section A and B which must not be saturated by 8 trains}
\vspace{-5 mm}
\end{figure*}

If the 8 trains are allowed to move freely, i.e., if their next endpoint is free, there is the possibility of creating deadlocks, i.e., a situation in which the 8 trains block each other in their expected progression. To solve this problem the scheduling algorithm of the ATS must take into consideration two \textit{critical sections} A and B -- i.e., zones of the layout in which a deadlock might occur -- which have the form of a ring of length 8 (see Fig.~\ref{fig:criticalregion}), and guarantee that these rings are never saturated with 8 trains -- further information on how critical sections are identified can be found in our previous work~\cite{mazzanti2014deadlock}. %, because these are precisely the sources of possible deadlocks. 
This can be modelled by using two global counters $RA$ and $RB$, which record the current number of trains inside these critical sections,
and by updating them whenever a train enters or exits these sections. For this purpose, each train mission $T_i$, with $i = 0 \ldots \text{MISSION\_LEN}$ (in our case MISSION\_LEN = 7)
, is associated with: a vector of increments/decrements $A_i$ to be applied to counter $RA$ at each step of progression; a vector $B_i$ of increments/decrements to be applied to counter $RB$.

For example, given $T_0=[1,9,10,13,15,20,23]$, and $A_0=[0, 0, 0, 1, 0,-1, 0]$, when \texttt{train0} moves from endpoint 10 to endpoint 13 ($P_0 = 3$) we must check that the +1 increment of $RA$ does not saturate the critical section A, i.e.,  $RA + A_0[P_0] \leq LA$ (in our case, $LA$ = 7); if the check passes then the train can proceed and safely update the counter $RA := RA + A_0[P_0]$. The maximum number of trains allowed in each critical section (i.e., $7$), will be expressed as $LA$ and $LB$ in the following. 

The models presented in Appendix A, which implement the algorithm described above, are deadlock-free, since the verification is being carried on as a final validation of a correct design. The actual possibility of having deadlocks, if the critical sections management were not supported or incorrectly implemented, can easily be observed by raising from 7 to 8 the values of the variables $LA$ or $LB$. 
%, i.e. allowing any train to freely advance if just its next endpoint is not occupied.

The case study presented here is actually a fragment of the complete TRACE-IT case study. In the original model the railway layout is much larger and the trains continually repeat cycling round missions. In that configuration further deadlocks situations may occur and further critical sections have to be defined and managed. The model considered in this case study represents just one of the three fragments in which the complete TRACE-IT layout has been decomposed to render the complexity of the problem amenable for formal verification. This is a typical procedure in the verification of real-world railway problems~\cite{winter2003modelling}.

The current design, in which each system state logically corresponds to a set of train progresses and each train movement logically corresponds to an atomic system evolution step, leads to a state-space of 1,636,535 configurations. This data is useful because it allows the user to cross-check the correctness of the encoding of this logical design in the various frameworks.

\section{Commonalities and Differences}
\label{sec:comparison}

\subsection{The basic blackboard model}

We want to build a model that describes all the possible evolutions of the system composed by the 8 trains, with purpose of verifying the correctness of the A0, ..., A7  and B0, ..., B7 tables that control the non saturation of the sections A and B, and the correctness of the assumption that the A and B sections are the only zones where a deadlock might occur.
The design skeleton we have in mind is that of a blackboard model, where a global space of common variables is read and updated by a set of atomic transformation operations. 
An atomic system evolution corresponds to a one-step movement of one train in the yard, which can occur when the next endpoint is free and when the move does not saturate neither the A section, nor the B section. 
We have encoded the above simple skeleton design using notations supported by 10 verification frameworks, namely UMC, Promela/SPIN, NuSMV/nuXmv, mCRL2, CPN Tools, FDR4, CADP, TLA+, UPPAAL and ProB. 
Within the context of this paper, our goal is to provide some feedback on the differences and traps that should be tackled when changing the reference frameworks, and the commonalities that would allow a simple translation from one framework to another.
Each framework surely has its own typical set of features that might lead to the best modelling and verification of a system, but, in this work, we are not interested in comparing the best way in which all the 10 frameworks could model the system. Instead, we are interested in seeing if, and to which extent, our basic design skeleton could be fitted with minimal transformations in all the frameworks taken into consideration.

In the following subsections we summarise some of the aspects that appear to characterise the differences of the various frameworks as evidenced by our specification problem. These observations can support the reader in making sense of the different models that are reported in Appendix A\footnote{In Appendix A we report solely the sequential cases -- according to the classification in Sect.~\ref{sec:structure} -- which are the most representative for our design.}, and attached to the current paper. 
More specifically, for each framework, we provide one or more model variants. The variants represent different modelling styles, according to the classification provided in Sect.~\ref{sec:structure}. 
In the case of CPN Tools, the different variants are associated to models with a different number of trains. Indeed, in our experiments, presented in~\cite{mazzanti2018towards,isola}, CPN Tools was not able to verify the case with eight trains, and models with a lower number or trains were tested. Table~\ref{tab:models} provides a brief description of the different variants considered, with the associated file names.

% \usepackage{multirow}
\begin{table}[]
\centering
\caption{Different models developed with associated frameworks.}
\label{tab:models}
\begin{footnotesize}
\begin{tabular}{|l|l|l|}
\hline
\textbf{Framework}                    & \textbf{File Name}                   & \textbf{Description}                         \\ \hline \hline
\multirow{2}{*}{\textbf{CADP}}   & cadp\_oneway8par.lnt                 & Parallel without shared memory               \\ \cline{2-3} 
                                 & cadp\_oneway8seq.lnt                 & Sequential                                   \\ \hline
\textbf{CPN Tools}               & cpn-oneway$<$X$>$.xml & Parallel without shared memory with X trains \\ \hline
\multirow{2}{*}{\textbf{FDR4}}   & fdr4\_oneway8par.txt                 & Parallel without shared memory               \\ \cline{2-3} 
                                 & fdr4\_oneway8seq.txt                 & Sequential                                   \\ \hline
\multirow{2}{*}{\textbf{mCRL2}}  & mcrl2\_oneway8par.txt                & Parallel without shared memory               \\ \cline{2-3} 
                                 & mcrl2\_oneway8seq.txt                & Sequential                                   \\ \hline
\textbf{ProB}                    & prob\_oneway8seq.mch                 & Sequential                                   \\ \hline
\textbf{NuSMV/nuXmv}             & smv\_oneway8-SM.smv                  & Sequential                                   \\ \hline
\textbf{SPIN}                    & spin\_oneway8.pml                    & Sequential                                   \\ \hline
\textbf{TLA+}                    & tla\_oneway8.txt                     & Sequential                                   \\ \hline
\textbf{UMC}                     & umc\_oneway8seq.txt                  & Sequential                                   \\ \hline
\multirow{2}{*}{\textbf{UPPAAL}} & uppaal-oneway8par.ta                 & Parallel with shared memory                  \\ \cline{2-3} 
                                 & uppaal-oneway8seq.ta                 & Sequential                                   \\ \hline
\end{tabular}
\end{footnotesize}
\end{table}

\subsection{System Design Structure}
\label{sec:structure}

The frameworks taken into account allow different kinds of model structures, which can be seen in our variants. 

\begin{description}
\item[Sequential]
With the sequential design structure the global system status is read and updated by a single sequential, nondeterministic process. This is the case that more directly reflects our initial design skeleton, and this structure has been modelled in all the considered frameworks\footnote{mcrl2\_oneway8seq.ta, cadp\_oneway8seq.lnt, fdr4\_oneway8seq.csp, umc\_oneway8seq.txt, spin\_oneway8.pml, prob\_oneway8.mch, tla_oneway8.txt, smv\_oneway8-SM.smv}, with the exception of CPN Tools. Indeed, this modelling style can be reproduced with CPN Tools, but it is not in line with the typical use of Petri Nets.
%since the Petri Nets formalism does not support this modelling style.

\item[Parallel without Shared Memory] With this design structure we indicate the case in which different parallel process interact among themselves in the absence of a common shared memory that could be directly read and updated by the processes. 
%In general, concurrent frameworks allow the user to design a system as a collection of interacting entities, but without allowing the presence of some shared memory that could be directly read and updated by the entities themselves. 
This is in general the case of concurrent frameworks, such as UMC, CPN, FDR4, CADP, mCRL2, where sets of processes (or a network layout in the case of Petri Nets) are used to model the system, and where a single entity might model the evolutions in time of a specific component of the system status (e.g., a variable). This is not our main reference scenario, however in the case of mCRL2, CADP, FDR4, CPN Tools we show alternative modelling examples that follow this design structure\footnote{mcrl2\_oneway8par.ta, cadp\_oneway8par.lnt, fdr4\_oneway8par.csp, cpn\_oneway8.xml, cpn\_oneway6-nocol.xml}.

\item[Parallel with Shared Memory]  
With this design structure a set of parallel processes share a common memory space, and, at the same time, may interact through inter-process communication. 
SPIN and UPPAAL are the only frameworks that allow the user to design a system in this way. An example of this model structure has been shown only in the case of UPPAAL\footnote{uppal-oneway8seq.ta}. 
\end {description}

subsection{Language Style}

Another evident difference among the various frameworks, is the overall style of the language used to specify the system.
For example, if we consider the way in which the transition relation (i.e., the system evolutions) are described, we can observe that three main approaches are followed by our considered frameworks. These three language styles can be qualified as \emph{imperative}, \emph{logical} and \emph{algebraic}, and are exemplified below with small fragments of code in the style of CADP-LNT~\cite{lnt}\footnote{ LTN is one of the languages supported by CADP, and is the language chosen for our experiment.},
%When referring to the language used, CADP is referred as CADP-LNT, since LNT is the language adopted in our case.}, 
TLA+ and FDR4, respectively.

\vspace{-3 mm}
\begin{alltt}
{\footnotesize 
if P0<6 then           (P0 < 6) &            System(P0,RA) =
   P0 := P0+1;         (P0' = P0+1) &          (P0 <6)  -> 
   RA := RA+A0[P0];    (RA' = RA+A0[P0+1])     System(P0+1,RA+A0[P0+1])
end;
}
\end{alltt}
\vspace{-3 mm}

In spite of the apparent difference, if the state transformation to be carried out during a system evolution is simple (like it happens in our case), the three styles are roughly equivalent, and translation from one style to the other can be performed with limited effort. 
%Indeed, in our case, the models developed 
%In the case of ProB imperative, algebraic (CSP from FDR4), logical notations (TLA+) (the three notations accepted by ProB) are internally translated into a prolog program.

\subsection{Arrays and Indexing}

In our example we do not have the need to use sophisticated data structures, and our design skeleton is just based on integer values and fixed-size tables of numbers.
Sometimes, e.g., in the case of UMC, SPIN, NuSMV/nuXmv, CADP-LNT, UPPAAL, TLA+, array-like types and indexing operations are natively supported by the specification language; other times, e.g., in the case of CPN Tools, FDR4, mCRL2, arrays should be represented as functions, or sequences, or lists, and the indexing operations possibly manually encoded as custom recursive functions. For example, in the case of FDR4 we have:

\vspace{-3 mm}
\begin{alltt}
{\footnotesize 
M0 = <1,9,10,13,15,20,23>  -- list of endpoint for the mission of train0

select_item(list,index) =    -- item selection, given an index 
  if index==0 then           --  (assuming index in the appropriate range)
     head(list) 
  else 
    select_item(tail(list),index-1);
}
\end{alltt}
\vspace{-3 mm}
  
\subsection{System Initialisation}

The different ways in which the frameworks treat system initialisation point out a difference that might trick an inexperienced designer.
Three different approaches can be recognised when a state variable in defined by the model, but not explicitly initialised at system startup.

\begin{description}
\item[Default Value]  
The uninitialised variables might get some default initial value (typically 0 for integers). This is the approach found in UMC, SPIN, UPPAAL.
\item[Error]   
The situation can be statically recognised as a design error, and notified to the designer. 
This is the approach followed by TLA+, ProB, CPN Tools, FDR4, mCRL2, CADP-LNT.
\item[Nondeterministic Assignment]  
The not explicitly initialised variable may nondeterministically get any of the possible values allowed by its type. This approach has been encountered only in in Nu\-SMV/\-nu\-Xmv. From one side this choice provides a powerful and flexible way to specify a rich set of possible system initial values, from the other side it might trick an inexperienced designer wrongly thinking that a classical default value (like 0) is used instead.
\end{description}
 
\subsection{The Transition Relation}

In all the considered  frameworks the transition relation is defined by rules that have the form:  \emph{guard-condition / state-transformation-effects}.
A possible question is what happens to the variables that are not explicitly modified by the \emph{state-transformation-effects}.
The situation is similar to the initialisation issue previously seen. Also in this case we have three different approaches:

\begin{description}
\item[Previous Value]   
The not explicitly assigned state variables preserve their previous value.
This is the approach followed by CPN, UPPAAL, FDR4, mCRL2, SPIN, UMC, ProB, CADP-LNT.
\item[Default Value]  
The not explicitly assigned state variables get a default \emph{null} value. This is what happens in the case of TLA+.
\item[Nondeterministic Assignment] 
The not explicitly assigned variable may nondeterministically get any of the possible values allowed by its type. This is what happens in the NuSMV/nuXmv case.
\end{description}

The difference among the three classes is evident if we compare the fragments of  \emph{state-transformation-effects} as they occur in CADP-LNT, TLA+ and NuSMV/nuXmv:

\vspace{-3 mm}
\begin{alltt}
{\footnotesize 
       P0 := P0+1;       (P0' = P0+1) &           next(P0) in P0+1 &
                         UNCHANGED<<P1,..P7>>     next(P1) in P1 &
                                                  ...
                                                  next(P7) in P7 
}
\end{alltt}
\vspace{-3 mm}

While with CADP-LNT it is not needed to make explicit that {\tt \footnotesize P1 ... P7} do not change their value, in TLA+ we need to use the keyword {\tt \footnotesize UNCHANGED}, 
and in NuSMV/nuXmv we have to explicitly state, for each variable, that the next value is equal to the one at the previous execution step.

Another relevant difference among the various frameworks is whether they allow the transition relation to be only \textit{partially} defined, i.e., 
are certain inputs and certain states allowed not to trigger a system evolution?

In our problem this situation actually occurs. For example, when a train cannot proceed because its next endpoint is occupied by another train, the rule describing the train progress cannot be applied. In all frameworks, with the exception of NuSMV/nuXmv, this does not represent a problem. It simply means that from such a system configuration state there is no outgoing edge corresponding to the movement of that train.

In the case of NuSMV/nuXmv instead the transition relation must be a \textit{total} function. This means that if a certain state configuration and a certain system input does not trigger an actual system evolution, we should equally explicitly state that the next system state is unchanged. If we fail to explicitly state that, the consequence is that the next state can become any state where all the system state variables nondeterministically get any of the values allowed by their type.
Notice that in this way we are introducing self loops in many states of the graph describing the system behaviour, and this has a certain impact on the way in which the system properties could be stated and verified. For example,  the user might be constrained to specify fairness constraints, or avoid the use of LTL formulas, or avoid CTL formulas like \emph{AF$<$statepredicate$>$}.
Indeed the verification approach of NuSMV always takes into consideration only infinite -- possibly fair, if requested -- traces\footnote{when using the \texttt{-bmc} option the behaviour might be different}.

\subsection{ Verification Techniques}

In our case the property we want to verify is that \textit{for all possible executions all the trains eventually complete their missions}. 
This property can be easily specified and verified in all the considered frameworks. 
However each framework provides original advanced verification features not supported by other frameworks. 
The possibility to translate a specification from a formalism to another might lead to several advantages:

\begin{itemize}
\item  We can increase the confidence of the verification results, given that none of the analysed frameworks are qualified at the highest integrity levels usually required by safety critical standards.
\item  We can exploit the specific strong points of more than one framework (e.g. the friendliness of a user interface, the ability to scale well, the possibility of generating program code or performing model based testing). 
\item We can verify a wider class of properties. For example, by importing a FDR4 model into ProB we can verify also LTL/CTL properties, by translating a model into UPPAAL we can introduce and verify further time related aspects, and so on. Table~\ref{tab:features} summarises the basic verification features that the considered frameworks make available.

\end{itemize}

\begin{table}[!h]
\centering
\caption{Verification features supported by the various frameworks}
\label{tab:features}
\begin{footnotesize}
\begin{tabular}{|c|l|}
\hline
\textbf{Framework}   & \textbf{Supported Verification Techniques}                       \\ \hline  \hline
\textbf{UMC}         & model checking CTL-like, state-event based logics                \\ \hline
\textbf{SPIN}        & model checking LTL, fairness requirements                        \\ \hline
\textbf{NuSMV/nuXMV} & LTL, CTL, PSL~\cite{psl}, SMT model checking, fairness requirements         \\ \hline
\textbf{CADP}        & MCL~\cite{mcl}, Parametric Mu-Calculus model checking, equivalence checking \\ \hline
\textbf{UPPAAL}      & MITL~\cite{mitl}, time-related, and probability related properties            \\ \hline
\textbf{TLA+}        & LTL, Theorem Proving, Proof Validations                          \\ \hline
\textbf{ProB}        & LTL, CTL model checking, constraints based checking  \\ \hline
\textbf{mCRL2}       & Parametric Mu Calculus model checking, equivalence checking \\ \hline
\textbf{FDR4}        & Refinement Checking, fairness requirements                       \\ \hline
\textbf{CPN}         & CTL, custom ML properties                                        \\ \hline
\end{tabular}
\end{footnotesize}
\end{table}

\subsection{Some Performance Data}

It is not a goal of our paper to make a comparative evaluation of the various frameworks in terms of scalability or performance. 
Nevertheless a summary of the  experienced times  when evaluating the property that \emph{for all possible executions all the trains eventually complete their missions} might still be a useful approximate indication of the impact of a certain system design approach / formal verification technique  in terms of performance.
The verification times presented in Table~\ref{tab:times} are expressed as ranges because they actually depend of the specific design approach adopted, on the specific formulas being evaluated, and on the specific options used during the tool execution. We refer to~\cite{mazzanti2018towards} for additional details.

\begin{table}[]
\centering
\caption{Indicative Summary of Evaluation Times}
\label{tab:times}
\begin{footnotesize}
\begin{tabular}{|c|l|}
\hline
\textbf{Framework}   & \textbf{Range of evalution times}        \\ \hline \hline
\textbf{UMC}         & 38 - 86 seconds                          \\ \hline
\textbf{SPIN}        & 13 - 47 seconds                          \\ \hline
\textbf{NuSMV/nuXMV} & 2.9 - 43 seconds                         \\ \hline
\textbf{CADP}        & 29 seconds                               \\ \hline
\textbf{UPPAAL}      & 16 seconds                               \\ \hline
\textbf{TLA+}        & 3 minutes                                \\ \hline
\textbf{ProB}        & 32 minutes                               \\ \hline
\textbf{mCRL2}       & 2 minutes -19 minutes                              \\ \hline
\textbf{FDR4}        & 15 seconds - 20 minutes                  \\ \hline
\textbf{CPN}         & unable to deal with the state-space size \\ \hline
\end{tabular}
\end{footnotesize}
\end{table}

\section{Conclusion}
\label{sec:conclusion}
The availability of CBTC systems relies on the existence of smart ATS systems that prevent the occurrence of deadlock situations in the metro network. 
In this paper, we present different models of a scheduling algorithm for an ATS, which was designed and verified to avoid deadlocks. 
Ten different formal frameworks are used, and different variants of system design structure are presented, according to the 
features made available by the frameworks. Differences in terms of language style, allowed data types, and treatments of the system evolution are observed, 
based on the developed models. In our future work, we plan to adapt our design to tools for model-based development such as Simulink/Stateflow, 
and SCADE, to explore their potential in terms of modelling styles and verification capabilities, and compare them with the other frameworks. 
Furthermore, in the context of the EU ASTRail project\footnote{\url{http://www.astrail.eu}} we are involved in a comparative analysis of formal and semi-formal 
tools in the railway domain. The experience gained with the different frameworks will be applied to provide diverse models for ERTMS/ETCS (European Rail Traffic Management System/European Train Control System) Level 3,
the next evolution of ERTMS/ETCS. This will allow us to further stress the capability of the frameworks with a different design, including time and probabilistic aspects. It shall be noticed that, in the current work, we did not discuss aspects related to the usability of the various frameworks. This issue is of paramount importance, as highlighted, among others, by Sirjani~\cite{rebeca}, and is also going to be considered in the context of the ASTRail project.  

\smallskip
\noindent
\textbf{Acknowledgements} 
This work has been partially funded by the ASTRail project. This project received funding from the Shift2Rail Joint Undertaking under the European Union’s Horizon 2020 research and innovation programme under grant agreement No 777561. The content of this paper reflects only the authors’ view and the Shift2Rail Joint Undertaking is not responsible for any use that may be made of the included information.

%\nocite{*}
\bibliographystyle{eptcs}
\bibliography{bibliography}
\section*{Appendix A}
\label{sec:models}
This appendix includes the sequential models for the different tools (when a textual representation is available). 
The all these models, together with the other graphical models for CPN Tools and ProB, can be retrieved from the MARS repository.

\subsection{CADP-LNT}
{\scriptsize 
\begin{verbatim}

module CADP_ONEWAY8SEQ is

------------------------------------------------------------------------------

type Train_Number is
   range 0 .. 7 of nat
end type

------------------------------------------------------------------------------

type Train_Mission is
    array [0 .. 6] of nat
end type

------------------------------------------------------------------------------

type Train_Constraint is
    array [0 .. 6] of int -- actually, of range -1 .. 1
end type

------------------------------------------------------------------------------

channel Movement is
   (Train : Train_Number)
end channel

------------------------------------------------------------------------------

process MAIN [MOVE : Movement, ARRIVED : none] is
   var P0, P1, P2, P3, P4, P5, P6, P7 : nat,
       RA, RB : int,
       LA, LB : int,
       T0, T1, T2, T3, T4, T5, T6, T7 : Train_Mission,
       A0, A1, A2, A3, A4, A5, A6, A7 : Train_Constraint,
       B0, B1, B2, B3, B4, B5, B6, B7 : Train_Constraint
   in
      P0 := 0;
      P1 := 0;
      P2 := 0;
      P3 := 0;
      P4 := 0;
      P5 := 0;
      P6 := 0;
      P7 := 0;
      RA := 1;
      RB := 1;
      LA := 7; -- limit for region A
      LB := 7; -- limit for region B
      -- ------------  train missions ------------
      T0 := Train_Mission ( 1, 9,10,13,15,20,23);
      T1 := Train_Mission ( 3, 9,10,13,15,20,24);
      T2 := Train_Mission ( 5,27,11,13,16,20,25);
      T3 := Train_Mission ( 7,27,11,13,16,20,26);
      T4 := Train_Mission (23,22,17,18,11, 9, 2);
      T5 := Train_Mission (24,22,17,18,11, 9, 4);
      T6 := Train_Mission (25,22,17,18,12,27, 6);
      T7 := Train_Mission (26,22,17,18,12,27, 8);
      -- -----------------------------------------
      
      -- ----- region A: train constraints ------
      A0 := Train_Constraint ( 0, 0, 0, 1, 0,-1, 0);
      A1 := Train_Constraint ( 0, 0, 0, 1, 0,-1, 0);
      A2 := Train_Constraint ( 0, 0, 1,-1, 0, 1, 0);
      A3 := Train_Constraint ( 0, 0, 1,-1, 0, 0, 0);
      A4 := Train_Constraint ( 0, 1, 0, 0,-1, 0, 0);
      A5 := Train_Constraint ( 0, 1, 0, 0,-1, 0, 0);
      A6 := Train_Constraint ( 0, 0, 0,-1, 0, 0, 0);
      A7 := Train_Constraint ( 0, 1, 0,-1, 0, 0, 0);
      -- -----------------------------------------
      
      -- ----- region B: train constraints ------
      B0 := Train_Constraint ( 0, 0, 0, 1, 0,-1, 0);
      B1 := Train_Constraint ( 0, 0, 0, 1, 0,-1, 0);
      B2 := Train_Constraint ( 0, 0, 1,-1, 0, 0, 0);
      B3 := Train_Constraint ( 0, 0, 1,-1, 0, 1, 0);
      B4 := Train_Constraint ( 0, 1, 0, 0,-1, 0, 0);
      B5 := Train_Constraint ( 0, 1, 0, 0,-1, 0, 0);
      B6 := Train_Constraint ( 0, 1, 0,-1, 0, 0, 0);
      B7 := Train_Constraint ( 0, 0, 0,-1, 0, 0, 0);
      -- -----------------------------------------


      loop
         select
            only if
              (P0 < 6) and
              (T0 [P0+1] != T1 [P1]) and -- next place of train0 not occupied by train1
              (T0 [P0+1] != T2 [P2]) and -- next place of train0 not occupied by train2
              (T0 [P0+1] != T3 [P3]) and
              (T0 [P0+1] != T4 [P4]) and
              (T0 [P0+1] != T5 [P5]) and
              (T0 [P0+1] != T6 [P6]) and
              (T0 [P0+1] != T7 [P7]) and -- next place of train0 not occupied by train7
              (RA + A0 [P0+1] <= LA) and -- progress of train0 does not saturate RA
              (RB + B0 [P0+1] <= LB)     -- progress of train0 does not saturate RD
            then
               MOVE (0 of Train_Number);
               P0 := P0 + 1;
               RA := RA + A0 [P0];
               RB := RB + B0 [P0]
            end if
          []
            only if
              (P1 < 6) and
              (T1 [P1+1] != T0 [P0]) and 
              (T1 [P1+1] != T2 [P2]) and
              (T1 [P1+1] != T3 [P3]) and
              (T1 [P1+1] != T4 [P4]) and
              (T1 [P1+1] != T5 [P5]) and
              (T1 [P1+1] != T6 [P6]) and
              (T1 [P1+1] != T7 [P7]) and 
              (RA + A1 [P1+1] <= LA) and
              (RB + B1 [P1+1] <= LB)    
            then
               MOVE (1 of Train_Number);
               P1 := P1 + 1;
               RA := RA + A1 [P1];
               RB := RB + B1 [P1]
            end if
          []
            only if
              (P2 < 6) and
              (T2 [P2+1] != T0 [P0]) and
              (T2 [P2+1] != T1 [P1]) and
              (T2 [P2+1] != T3 [P3]) and
              (T2 [P2+1] != T4 [P4]) and
              (T2 [P2+1] != T5 [P5]) and
              (T2 [P2+1] != T6 [P6]) and
              (T2 [P2+1] != T7 [P7]) and  
              (RA + A2 [P2+1] <= LA) and
              (RB + B2 [P2+1] <= LB)     
            then
               MOVE (2 of Train_Number);
               P2 := P2 + 1;
               --if ( P2 == 13 ) then P2 := 0 end if; 
               RA := RA + A2 [P2];
               RB := RB + B2 [P2]
            end if
          []
            only if
              (P3 < 6) and
              (T3 [P3+1] != T0 [P0])  and
              (T3 [P3+1] != T1 [P1])  and
              (T3 [P3+1] != T2 [P2])  and
              (T3 [P3+1] != T4 [P4])  and
              (T3 [P3+1] != T5 [P5])  and
              (T3 [P3+1] != T6 [P6])  and
              (T3 [P3+1] != T7 [P7])  and
              (RA + A3 [P3+1] <= LA) and
              (RB + B3 [P3+1] <= LB)  
            then
               MOVE (3 of Train_Number);
               P3 := P3 + 1;
               RA := RA + A3 [P3];
               RB := RB + B3 [P3]
            end if
          []
            only if
              (P4 < 6) and
              (T4 [P4+1] != T0 [P0])  and
              (T4 [P4+1] != T1 [P1])  and
              (T4 [P4+1] != T2 [P2])  and
              (T4 [P4+1] != T3 [P3])  and
              (T4 [P4+1] != T5 [P5])  and
              (T4 [P4+1] != T6 [P6])  and
              (T4 [P4+1] != T7 [P7])  and
              (RA + A4 [P4+1] <= LA) and    
              (RB + B4 [P4+1] <= LB) 
            then
               MOVE (4 of Train_Number);
               P4 := P4 + 1;
               RA := RA + A4 [P4];
               RB := RB + B4 [P4]
            end if
          []
            only if
              (P5 < 6) and
              (T5 [P5+1] != T0 [P0]) and
              (T5 [P5+1] != T1 [P1]) and 
              (T5 [P5+1] != T2 [P2]) and
              (T5 [P5+1] != T3 [P3]) and
              (T5 [P5+1] != T4 [P4]) and
              (T5 [P5+1] != T6 [P6]) and
              (T5 [P5+1] != T7 [P7]) and
              (RA + A5 [P5+1] <= LA) and
              (RB + B5 [P5+1] <= LB)    
            then
               MOVE (5 of Train_Number);
               P5 := P5 + 1;
               RA := RA + A5 [P5];
               RB := RB + B5 [P5]
            end if
          []
            only if
              (P6 < 6) and
              (T6 [P6+1] != T0 [P0]) and
              (T6 [P6+1] != T1 [P1]) and
              (T6 [P6+1] != T2 [P2]) and
              (T6 [P6+1] != T3 [P3]) and
              (T6 [P6+1] != T4 [P4]) and
              (T6 [P6+1] != T5 [P5]) and
              (T6 [P6+1] != T7 [P7]) and
              (RA + A6 [P6+1] <= LA) and
              (RB + B6 [P6+1] <= LB)
            then
               MOVE (6 of Train_Number);
               P6 := P6 + 1;
               RA := RA + A6 [P6];
               RB := RB + B6 [P6]
            end if
          []
            only if
              (P7 < 6) and
              (T7 [P7+1] != T0 [P0])  and
              (T7 [P7+1] != T1 [P1])  and
              (T7 [P7+1] != T2 [P2])  and
              (T7 [P7+1] != T3 [P3])  and
              (T7 [P7+1] != T4 [P4])  and
              (T7 [P7+1] != T5 [P5])  and
              (T7 [P7+1] != T6 [P6])  and
              (RA + A7 [P7+1] <= LA) and
              (RB + B7 [P7+1] <= LB)
            then
               MOVE (7 of Train_Number);
               P7 := P7 + 1;
               RA := RA + A7 [P7];
               RB := RB + B7 [P7]
            end if
          []
            -- ALL TRAINS RUNNING
            only if (P0 == 6) and (P1 == 6) and (P2 == 6) and (P3 == 6) and 
                    (P4 == 6) and (P5 == 6) and (P6 == 6) and (P7 == 6) 
            then
               ARRIVED
            end if
         end select
      end loop
   end var
end process

end module

--
-- lnt.open cadp_oneway8.lnt generator x
-- bcg_info x.bcg
--     
--  1_636_545 states
--  7_134_233 transitions
--     
-- time lnt.open cadp_oneway8small.lnt evaluator4 cadpafarr.mcl
--      cadpafarr.mcl ==  mu XXX.(([not ARRIVED] XXX) and (<true> true))
--      cadpafarr.mcl ==  [ true* ] < true* . ARRIVED > true
--      cadpafarr.mcl ==  [ true* ] < true> true
--
-- >  TRUE
-- > 
-- > real	0m29.648s
-- > user	0m28.341s
-- > sys	0m1.078s
-- Evaluator4 Memory 78MB
--

\end{verbatim}
}

\subsection{FDR4}

{\scriptsize
\begin{verbatim}


M0 = < 1, 9,10,13,15,20,23>
M1 = < 3, 9,10,13,15,20,24>
M2 = < 5,27,11,13,16,20,25>
M3 = < 7,27,11,13,16,20,26>
M4 = <23,22,17,18,11, 9, 2>
M5 = <24,22,17,18,11, 9, 4>
M6 = <25,22,17,18,12,27, 6>
M7 = <26,22,17,18,12,27, 8>


  ------ region A: train constraints ------
A0 = < 0, 0, 0, 1, 0,-1, 0> -- G1
A1 = < 0, 0, 0, 1, 0,-1, 0> -- R1
A2 = < 0, 0, 1,-1, 0, 1, 0> -- Y1
A3 = < 0, 0, 1,-1, 0, 0, 0> -- B1
A4 = < 0, 1, 0, 0,-1, 0, 0> -- G2
A5 = < 0, 1, 0, 0,-1, 0, 0> -- R2
A6 = < 0, 0, 0,-1, 0, 0, 0> -- Y2
A7 = < 0, 1, 0,-1, 0, 0, 0> -- B2 
 ------------------------------------------

  ------- region B: train constraints ------
B0 = < 0, 0, 0, 1, 0,-1, 0> -- G1
B1 = < 0, 0, 0, 1, 0,-1, 0> -- R1
B2 = < 0, 0, 1,-1, 0, 0, 0> -- Y1
B3 = < 0, 0, 1,-1, 0, 1, 0> -- B1
B4 = < 0, 1, 0, 0,-1, 0, 0> -- G2
B5 = < 0, 1, 0, 0,-1, 0, 0> -- R2
B6 = < 0, 1, 0,-1, 0, 0, 0> -- Y2
B7 = < 0, 0, 0,-1, 0, 0, 0> -- B2
  ------------------------------------------
LA = 7
LB = 7

el(y,x) = if x==0 then head(y) else el(tail(y),x-1)

--channel move:{1..27}.{1..27}.{ -1..1}.{ -1..1}
channel move
channel  arrived

AllTrains (P0, P1, P2, P3, P4, P5, P6, P7, RA, RB) = 
     (P0 < 6 and    --  train0 has not yet reached all the steps of its mission
      el(T0,P0+1) != el(T1,P1)  and -- next place of train0 not occupied by train1
      el(T0,P0+1) != el(T2,P2)  and -- next place of train0 not occupied by train2
      el(T0,P0+1) != el(T3,P3)  and
      el(T0,P0+1) != el(T4,P4)  and
      el(T0,P0+1) != el(T5,P5)  and
      el(T0,P0+1) != el(T6,P6)  and
      el(T0,P0+1) != el(T7,P7)  and -- next place of train0 not occupied by train7
      RA + el(A0,P0+1) <= LA and  -- progress of train0 does not saturate RA 
      RB + el(B0,P0+1) <= LB     -- progress of train0 does not saturate RB
     )  &
       move -> AllTrains(P0+1,P1,P2,P3,P4,P5,P6,P7,RA+el(A0,P0+1),RB+el(B0,P0+1))
     []
     (P1 < 6 and
      el(T1,P1+1) != el(T0,P0)  and
      el(T1,P1+1) != el(T2,P2)  and
      el(T1,P1+1) != el(T3,P3)  and
      el(T1,P1+1) != el(T4,P4)  and
      el(T1,P1+1) != el(T5,P5)  and
      el(T1,P1+1) != el(T6,P6)  and
      el(T1,P1+1) != el(T7,P7)  and
      RA + el(A1,P1+1) <= LA and    
      RB + el(B1,P1+1) <= LB
     )  &
       move -> AllTrains(P0,P1+1,P2,P3,P4,P5,P6,P7,RA+el(A1,P1+1),RB+el(B1,P1+1))
     []
     (P2 < 6 and
      el(T2,P2+1) != el(T0,P0)  and
      el(T2,P2+1) != el(T1,P1)  and
      el(T2,P2+1) != el(T3,P3)  and
      el(T2,P2+1) != el(T4,P4)  and
      el(T2,P2+1) != el(T5,P5)  and
      el(T2,P2+1) != el(T6,P6)  and
      el(T2,P2+1) != el(T7,P7)  and
      RA + el(A2,P2+1) <= LA and    
      RB + el(B2,P2+1) <= LB
     )  &
       move -> AllTrains(P0,P1,P2+1,P3,P4,P5,P6,P7,RA+el(A2,P2+1),RB+el(B2,P2+1))
     []
     (P3 < 6 and
      el(T3,P3+1) != el(T0,P0)  and
      el(T3,P3+1) != el(T1,P1)  and
      el(T3,P3+1) != el(T2,P2)  and
      el(T3,P3+1) != el(T4,P4)  and
      el(T3,P3+1) != el(T5,P5)  and
      el(T3,P3+1) != el(T6,P6)  and
      el(T3,P3+1) != el(T7,P7)  and
      RA + el(A3,P3+1) <= LA and    
      RB + el(B3,P3+1) <= LB
     )  &
       move -> AllTrains(P0,P1,P2,P3+1,P4,P5,P6,P7,RA+el(A3,P3+1),RB+el(B3,P3+1))
     []
     (P4 < 6 and
      el(T4,P4+1) != el(T0,P0)  and
      el(T4,P4+1) != el(T1,P1)  and
      el(T4,P4+1) != el(T2,P2)  and
      el(T4,P4+1) != el(T3,P3)  and
      el(T4,P4+1) != el(T5,P5)  and
      el(T4,P4+1) != el(T6,P6)  and
      el(T4,P4+1) != el(T7,P7)  and
      RA + el(A4,P4+1) <= LA and    
      RB + el(B4,P4+1) <= LB
     )  &
       move -> AllTrains(P0,P1,P2,P3,P4+1,P5,P6,P7,RA+el(A4,P4+1),RB+el(B4,P4+1))
     []
     (P5 < 6 and
      el(T5,P5+1) != el(T0,P0)  and
      el(T5,P5+1) != el(T1,P1)  and
      el(T5,P5+1) != el(T2,P2)  and
      el(T5,P5+1) != el(T3,P3)  and
      el(T5,P5+1) != el(T4,P4)  and
      el(T5,P5+1) != el(T6,P6)  and
      el(T5,P5+1) != el(T7,P7)  and
      RA + el(A5,P5+1) <= LA and    
      RB + el(B5,P5+1) <= LB
     )  &
       move -> AllTrains(P0,P1,P2,P3,P4,P5+1,P6,P7,RA+el(A5,P5+1),RB+el(B5,P5+1))
     []
     (P6 < 6 and
      el(T6,P6+1) != el(T0,P0)  and
      el(T6,P6+1) != el(T1,P1)  and
      el(T6,P6+1) != el(T2,P2)  and
      el(T6,P6+1) != el(T3,P3)  and
      el(T6,P6+1) != el(T4,P4)  and
      el(T6,P6+1) != el(T5,P5)  and
      el(T6,P6+1) != el(T7,P7)  and
      RA + el(A6,P6+1) <= LA and    
      RB + el(B6,P6+1) <= LB
     )  &
       move -> AllTrains(P0,P1,P2,P3,P4,P5,P6+1,P7,RA+el(A6,P6+1),RB+el(B6,P6+1))
     []
     (P7 < 6 and
      el(T7,P7+1) != el(T0,P0)  and
      el(T7,P7+1) != el(T1,P1)  and
      el(T7,P7+1) != el(T2,P2)  and
      el(T7,P7+1) != el(T3,P3)  and
      el(T7,P7+1) != el(T4,P4)  and
      el(T7,P7+1) != el(T5,P5)  and
      el(T7,P7+1) != el(T6,P6)  and
      RA + el(A7,P7+1) <= LA and    
      RB + el(B7,P7+1) <= LB
     )  &
       move -> AllTrains(P0,P1,P2,P3,P4,P5,P6,P7+1,RA+el(A7,P7+1),RB+el(B7,P7+1))
     []
     ((P0 == 6) and (P1 ==6) and (P2 ==6) and (P3 ==6)  and 
      (P4 ==6) and (P5 ==6) and (P6 ==6) and (P7 ==6)
     ) &
       arrived -> STOP

--------------------------
ASYS = AllTrains(0,0,0,0,0,0,0,0, 1,1)\{move}
--------------------------
--  compression is helpful for two verifications/visualization
-- NSYS = normal(ASYS)
-- assert SPEC [FD= NSYS
--------------------------


--------------------------
SPEC = arrived -> STOP
--------------------------

assert SPEC [FD= ASYS

-- -------- verfication process : ---------------
--  time refines --refinement-storage-file-path swapdir fdr4_oneway8seq.txt
--



\end{verbatim}
}

\section{mCRL2}

{\scriptsize
\begin{verbatim}

%--------------------------------------------------------------
% T0 := [ 1, 9,10,13,15,20,23,22]; -- G1
% T1 := [ 3, 9,10,13,15,20,24,22]; -- R1
% T2 := [ 5,27,11,13,16,20,25,22]; -- Y1
% T3 := [ 7,27,11,13,16,20,26,22]; -- B1
% T4 := [23,22,17,18,11, 9, 2, 1]; -- G2
% T5 := [24,22,17,18,11, 9, 4, 3]; -- R2
% T6 := [25,22,17,18,12,27, 6, 5]; -- Y2
% T7 := [26,22,17,18,12,27, 8, 7]; -- B2
%--------------------------------------------------------------

map T0: Nat -> Nat;
 eqn T0(0)= 1; T0(1)= 9; T0(2)=10; T0( 3)=13; T0( 4)=15; T0( 5)=20; T0( 6)=23;
     
map T1: Nat -> Nat;
 eqn T1(0)= 3; T1(1)=9;  T1(2)=10; T1( 3)=13; T1( 4)=15; T1( 5)=20; T1( 6)=24;
  
map T2: Nat -> Nat;
 eqn T2(0)= 5; T2(1)=27; T2(2)=11; T2( 3)=13; T2( 4)=16; T2( 5)=20; T2( 6)=25;
  
map T3: Nat -> Nat;
 eqn T3(0)= 7; T3(1)=27; T3(2)=11; T3( 3)=13; T3( 4)=16; T3( 5)=20; T3( 6)=26;
  
map T4: Nat -> Nat;
 eqn T4(0)=23; T4(1)=22; T4(2)=17; T4( 3)=18; T4( 4)=11; T4( 5)= 9; T4( 6)= 2;
  
map T5: Nat -> Nat;
 eqn T5(0)=24; T5(1)=22; T5(2)=17; T5( 3)=18; T5( 4)=11; T5( 5)=9;  T5( 6)= 4;
  
map T6: Nat -> Nat;
 eqn T6(0)=25; T6(1)=22; T6(2)=17; T6(3)=18; T6(4)=12; T6(5)=27; T6(6)=6;
     T6(7)= 5; T6(8)=27; T6(9)=11; T6(10)=13; T6(11)=16; T6(12)=20; T6(13)=25;
     
map T7: Nat -> Nat;
 eqn T7(0)=26; T7(1)=22; T7(2)=17; T7( 3)=18; T7( 4)=12; T7( 5)=27; T7( 6)= 8;
  
% ------ region A: train constraints ------
%         0  1  2  3  4  5  6
% A0 := [ 0, 0, 0, 1, 0,-1, 0]; -- G1
% A1 := [ 0, 0, 0, 1, 0,-1, 0]; -- R1
% A2 := [ 0, 0, 1,-1, 0, 1, 0]; -- Y1
% A3 := [ 0, 0, 1,-1, 0, 0, 0]; -- B1
% A4 := [ 0, 1, 0, 0,-1, 0, 0]; -- G2
% A5 := [ 0, 1, 0, 0,-1, 0, 0]; -- R2
% A6 := [ 0, 0, 0,-1, 0, 0, 0]; -- Y2
% A7 := [ 0, 1, 0,-1, 0, 0, 0]; -- B2
% ------------------------------------------
map  LA: Nat;  %  limit for region A
 eqn  LA = 7;
 
map A0: Nat -> Int;
 eqn A0(0)=0; A0(1)=0; A0(2)=0; A0( 3)= 1; A0( 4)=0; A0( 5)=-1; A0( 6)=0;
 
map A1: Nat -> Int;
 eqn A1(0)=0; A1(1)=0; A1(2)=0; A1( 3)= 1; A1( 4)=0; A1( 5)=-1; A1( 6)=0;
 
map A2: Nat -> Int;
 eqn A2(0)=0; A2(1)=0; A2(2)= 1; A2( 3)=-1; A2( 4)=0; A2( 5)= 1; A2( 6)=0;
 
map A3: Nat -> Int;
 eqn A3(0)=0; A3(1)=0; A3(2)= 1; A3( 3)=-1; A3( 4)=0; A3( 5)= 0; A3( 6)=0;
 
map A4: Nat -> Int;
 eqn A4(0)=0; A4(1)=1; A4(2)=0; A4( 3)=0; A4( 4)=-1; A4( 5)= 0; A4( 6)=0;
 
map A5: Nat -> Int;
 eqn A5(0)=0; A5(1)=1; A5(2)=0; A5( 3)=0; A5( 4)=-1; A5( 5)= 0; A5( 6)=0;
 
map A6: Nat -> Int;
 eqn A6(0)=0; A6(1)=0; A6(2)=0; A6( 3)=-1; A6( 4)=0; A6( 5)= 0; A6( 6)=0;
 
map A7: Nat -> Int;
 eqn A7(0)=0; A7(1)=1; A7(2)=0; A7( 3)=-1; A7( 4)=0; A7( 5)= 0; A7( 6)=0;

% ------- region B: train constraints ------
%         0  1  2  3  4  5  6
% B0 := [ 0, 0, 0, 1, 0,-1, 0]; -- G1
% B1 := [ 0, 0, 0, 1, 0,-1, 0]; -- R1
% B2 := [ 0, 0, 1,-1, 0, 0, 0]; -- Y1
% B3 := [ 0, 0, 1,-1, 0, 1, 0]; -- B1
% B4 := [ 0, 1, 0, 0,-1, 0, 0]; -- G2
% B5 := [ 0, 1, 0, 0,-1, 0, 0]; -- R2
% B6 := [ 0, 1, 0,-1, 0, 0, 0]; -- Y2
% B7 := [ 0, 0, 0,-1, 0, 0, 0]; -- B2
% ------------------------------------------
map  LB: Nat;  %  limit for region B
 eqn  LB = 7;
 
map B0: Nat -> Int;
 eqn B0(0)=0; B0(1)=0; B0(2)=0; B0( 3)= 1; B0( 4)=0; B0( 5)=-1; B0( 6)=0;
 
map B1: Nat -> Int;
 eqn B1(0)=0; B1(1)=0; B1(2)=0; B1( 3)= 1; B1( 4)=0; B1( 5)=-1; B1( 6)=0;
 
map B2: Nat -> Int;
 eqn B2(0)=0; B2(1)=0; B2(2)= 1; B2( 3)=-1; B2( 4)=0; B2( 5)= 0; B2( 6)=0;
 
map B3: Nat -> Int;
 eqn B3(0)=0; B3(1)=0; B3(2)= 1; B3( 3)=-1; B3( 4)=0; B3( 5)= 1; B3( 6)=0;
 
map B4: Nat -> Int;
 eqn B4(0)=0; B4(1)=1; B4(2)=0; B4( 3)=-0; B4( 4)=-1; B4( 5)= 0; B4( 6)=0;
 
map B5: Nat -> Int;
 eqn B5(0)=0; B5(1)=1; B5(2)=0; B5( 3)=0; B5( 4)=-1; B5( 5)= 0; B5( 6)=0;
 
map B6: Nat -> Int;
 eqn B6(0)=0; B6(1)=1; B6(2)=0; B6( 3)=-1; B6( 4)=0; B6( 5)= 0; B6( 6)=0;
 
map B7: Nat -> Int;
 eqn B7(0)=0; B7(1)=0; B7(2)=0; B7( 3)=-1; B7( 4)=0; B7( 5)= 0; B7( 6)=0;

act  arrived;
     move: Nat;

proc AllTrains(
   P0:Nat,P1:Nat,P2:Nat,P3:Nat,P4:Nat,P5:Nat,P6:Nat,P7:Nat,RA: Int,RB: Int) = 
     (P0 < 6 &&    %  train0 has not yet reached all the steps of its mission
      T0(P0+1) != T1(P1)  && % next place of train0 not occupied by train1
      T0(P0+1) != T2(P2)  && % next place of train0 not occupied by train2
      T0(P0+1) != T3(P3)  &&
      T0(P0+1) != T4(P4)  &&
      T0(P0+1) != T5(P5)  &&
      T0(P0+1) != T6(P6)  &&
      T0(P0+1) != T7(P7)  && % next place of train0 not occupied by train7
      RA + A0(P0+1) <= LA &&  % progress of train0 does not saturate RA 
      RB + B0(P0+1) <= LB     % progress of train0 does not saturate RB
     )  ->
       move(0). AllTrains(P0+1,P1,P2,P3,P4,P5,P6,P7,RA+A0(P0+1),RB+B0(P0+1))
     +
     (P1 < 6 &&
      T1(P1+1) != T0(P0)  &&
      T1(P1+1) != T2(P2)  &&
      T1(P1+1) != T3(P3)  &&
      T1(P1+1) != T4(P4)  &&
      T1(P1+1) != T5(P5)  &&
      T1(P1+1) != T6(P6)  &&
      T1(P1+1) != T7(P7)  &&
      RA + A1(P1+1) <= LA &&    
      RB + B1(P1+1) <= LB
     )  ->
       move(1). AllTrains(P0,P1+1,P2,P3,P4,P5,P6,P7,RA+A1(P1+1),RB+B1(P1+1))
     +
     (P2 < 6 &&
      T2(P2+1) != T0(P0)  &&
      T2(P2+1) != T1(P1)  &&
      T2(P2+1) != T3(P3)  &&
      T2(P2+1) != T4(P4)  &&
      T2(P2+1) != T5(P5)  &&
      T2(P2+1) != T6(P6)  &&
      T2(P2+1) != T7(P7)  &&
      RA + A2(P2+1) <= LA &&    
      RB + B2(P2+1) <= LB
     )  ->
       move(2). AllTrains(P0,P1,P2+1,P3,P4,P5,P6,P7,RA+A2(P2+1),RB+B2(P2+1))
     +
     (P3 < 6 &&
      T3(P3+1) != T0(P0)  &&
      T3(P3+1) != T1(P1)  &&
      T3(P3+1) != T2(P2)  &&
      T3(P3+1) != T4(P4)  &&
      T3(P3+1) != T5(P5)  &&
      T3(P3+1) != T6(P6)  &&
      T3(P3+1) != T7(P7)  &&
      RA + A3(P3+1) <= LA &&    
      RB + B3(P3+1) <= LB
     )  ->
       move(3). AllTrains(P0,P1,P2,P3+1,P4,P5,P6,P7,RA+A3(P3+1),RB+B3(P3+1))
     +
     (P4 < 6 &&
      T4(P4+1) != T0(P0)  &&
      T4(P4+1) != T1(P1)  &&
      T4(P4+1) != T2(P2)  &&
      T4(P4+1) != T3(P3)  &&
      T4(P4+1) != T5(P5)  &&
      T4(P4+1) != T6(P6)  &&
      T4(P4+1) != T7(P7)  &&
      RA + A4(P4+1) <= LA &&    
      RB + B4(P4+1) <= LB
     )  ->
       move(4). AllTrains(P0,P1,P2,P3,P4+1,P5,P6,P7,RA+A4(P4+1),RB+B4(P4+1))
     +
     (P5 < 6 &&
      T5(P5+1) != T0(P0)  &&
      T5(P5+1) != T1(P1)  &&
      T5(P5+1) != T2(P2)  &&
      T5(P5+1) != T3(P3)  &&
      T5(P5+1) != T4(P4)  &&
      T5(P5+1) != T6(P6)  &&
      T5(P5+1) != T7(P7)  &&
      RA + A5(P5+1) <= LA &&    
      RB + B5(P5+1) <= LB
     )  ->
       move(5). AllTrains(P0,P1,P2,P3,P4,P5+1,P6,P7,RA+A5(P5+1),RB+B5(P5+1))
     +
     (P6 < 6 &&
      T6(P6+1) != T0(P0)  &&
      T6(P6+1) != T1(P1)  &&
      T6(P6+1) != T2(P2)  &&
      T6(P6+1) != T3(P3)  &&
      T6(P6+1) != T4(P4)  &&
      T6(P6+1) != T5(P5)  &&
      T6(P6+1) != T7(P7)  &&
      RA + A6(P6+1) <= LA &&    
      RB + B6(P6+1) <= LB
     )  ->
       move(6). AllTrains(P0,P1,P2,P3,P4,P5,P6+1,P7,RA+A6(P6+1),RB+B6(P6+1))
     +
     (P7 < 6 &&
      T7(P7+1) != T0(P0)  &&
      T7(P7+1) != T1(P1)  &&
      T7(P7+1) != T2(P2)  &&
      T7(P7+1) != T3(P3)  &&
      T7(P7+1) != T4(P4)  &&
      T7(P7+1) != T5(P5)  &&
      T7(P7+1) != T6(P6)  &&
      RA + A7(P7+1) <= LA &&    
      RB + B7(P7+1) <= LB
     )  ->
       move(7). AllTrains(P0,P1,P2,P3,P4,P5,P6,P7+1,RA+A7(P7+1),RB+B7(P7+1))

     +
     ((P0 ==6) && (P1 ==6) && (P2 ==6) && (P3 ==6)  && 
      (P4 ==6) && (P5 ==6) && (P6 ==6) && (P7 ==6)) ->
       arrived . AllTrains(P0,P1,P2,P3,P4,P5,P6,P7,RA,RB);

init   AllTrains(0,0,0,0,0,0,0,0, 1,1);
     

%%%%%%%%%%% verfication process : %%%%%%%%%%%%%%%%%
% 
% mcrl22lps mcrl2_oneway8seq.txt temp.lps
% lps2pbes -fformula.mcf  temp.lps  temp.pbes
%     formula.mcf=  "mu X.(([!arrived]X) && (<true> true))"  ==  AF {arrived}
% time pbes2bool -s2 -vrjittyc temp.pbes
% > real	1m53.996s
% > user	1m52.243s
% > sys	    0m1.583s
% > USED VIRTUAL MEMORY  1.06 GB
%
% time pbes2bool -s2 temp.pbes
% > real  19m47.753s
%
% lps2lts -rjittyc --verbose temp.lps temp.lts 
% ltsinfo temp.lts
% >  state space stats: (49 levels, 1_636_545 states and 7_134_233 transitions)
% >
% >
%
% formula.mcf=  "mu X.(([!arrived]X) && (<true> true))"  ==  AF {arrived}
% formula.mcf=  "[ true* ] < true* . ARRIVED > true"     ==  AG EF {arrived}
%  pbes2bool  -c print evidence on validity of formula, nedded by lpsxsim for 
% 
% lps2lts --verbose –rjittyc –cached -D -F  temp.lps temp.lts
%    ( -sdepth  -sBreadth --suppress --todo=maxnum --no-info
%
% the generation of counter-examples
%  pbes2bool -zb    breadth-first
%  pbes2bool -zd    depth-first
%  pbes2bool --todo  -strategy  
%%%%%%%%%%%%%%%%%%%%%%%%%%%%%%%%%%%%%%%%%%%%%%%%%%%%%


\end{verbatim}
}

\section{ProB}

{\scriptsize
\begin{verbatim}
/* https://www3.hhu.de/stups/prob/index.php/Summary_of_B_Syntax */

MACHINE Oneway
 
DEFINITIONS
  SET_PREF_MAXINT == 1;
  SET_PREF_MININT == 0
  
CONSTANTS  
   T0,T1,T2,T3,T4,T5,T6,T7,
   A0,A1,A2,A3,A4,A5,A6,A7,
   B0,B1,B2,B3,B4,B5,B6,B7,
   LA,LB

PROPERTIES 
/*
------- train missions --------------
T0: int[] := [ 1, 9,10,13,15,20,23]; -- G1
T1: int[] := [ 3, 9,10,13,15,20,24]; -- R1
T2: int[] := [ 5,27,11,13,16,20,25]; -- Y1
T3: int[] := [ 7,27,11,13,16,20,26]; -- B1
T4: int[] := [23,22,17,18,11, 9, 2]; -- G2
T5: int[] := [24,22,17,18,11, 9, 4]; -- R2
T6: int[] := [25,22,17,18,12,27, 6]; -- Y2
T7: int[] := [26,22,17,18,12,27, 8]; -- B2

------ region A: train constraints ------
A0: int[] := [ 0, 0, 0, 1, 0,-1, 0]; -- G1
A1: int[] := [ 0, 0, 0, 1, 0,-1, 0]; -- R1
A2: int[] := [ 0, 0, 1,-1, 0, 1, 0]; -- Y1
A3: int[] := [ 0, 0, 1,-1, 0, 0, 0]; -- B1
A4: int[] := [ 0, 1, 0, 0,-1, 0, 0]; -- G2
A5: int[] := [ 0, 1, 0, 0,-1, 0, 0]; -- R2
A6: int[] := [ 0, 0, 0,-1, 0, 0, 0]; -- Y2
A7: int[] := [ 0, 1, 0,-1, 0, 0, 0]; -- B2

------- region B: train constraints ------
B0: int[] := [ 0, 0, 0, 1, 0,-1, 0]; -- G1
B1: int[] := [ 0, 0, 0, 1, 0,-1, 0]; -- R1
B2: int[] := [ 0, 0, 1,-1, 0, 0, 0]; -- Y1
B3: int[] := [ 0, 0, 1,-1, 0, 1, 0]; -- B1
B4: int[] := [ 0, 1, 0, 0,-1, 0, 0]; -- G2
B5: int[] := [ 0, 1, 0, 0,-1, 0, 0]; -- R2
B6: int[] := [ 0, 1, 0,-1, 0, 0, 0]; -- Y2
B7: int[] := [ 0, 0, 0,-1, 0, 0, 0]; -- B2
------------------------------------------
*/

  T0 : 0..6 --> 1..27 & 
  T1 : 0..6 --> 1..27 & 
  T2 : 0..6 --> 1..27 & 
  T3 : 0..6 --> 1..27 & 
  T4 : 0..6 --> 1..27 & 
  T5 : 0..6 --> 1..27 & 
  T6 : 0..6 --> 1..27 &
  T7 : 0..6 --> 1..27 &
  
  A0 : 0..6 --> -1..1 &
  A1 : 0..6 --> -1..1 & 
  A2 : 0..6 --> -1..1 & 
  A3 : 0..6 --> -1..1 & 
  A4 : 0..6 --> -1..1 & 
  A5 : 0..6 --> -1..1 & 
  A6 : 0..6 --> -1..1 &
  A7 : 0..6 --> -1..1 &
  
  B0 : 0..6 --> -1..1 &
  B1 : 0..6 --> -1..1 & 
  B2 : 0..6 --> -1..1 & 
  B3 : 0..6 --> -1..1 & 
  B4 : 0..6 --> -1..1 & 
  B5 : 0..6 --> -1..1 & 
  B6 : 0..6 --> -1..1 &
  B7 : 0..6 --> -1..1 &
  
  T0(0)= 1 & T0(1)= 9 & T0(2)=10 & T0(3)=13 & T0(4)=15 & T0(5)=20 & T0(6)=23 &
  T1(0)= 3 & T1(1)= 9 & T1(2)=10 & T1(3)=13 & T1(4)=15 & T1(5)=20 & T1(6)=24 &
  T2(0)= 5 & T2(1)=27 & T2(2)=11 & T2(3)=13 & T2(4)=16 & T2(5)=20 & T2(6)=25 &
  T3(0)= 7 & T3(1)=27 & T3(2)=11 & T3(3)=13 & T3(4)=16 & T3(5)=20 & T3(6)=26 &
  T4(0)=23 & T4(1)=22 & T4(2)=17 & T4(3)=18 & T4(4)=11 & T4(5)= 9 & T4(6)=2 &
  T5(0)=24 & T5(1)=22 & T5(2)=17 & T5(3)=18 & T5(4)=11 & T5(5)= 9 & T5(6)=4 &
  T6(0)=25 & T6(1)=22 & T6(2)=17 & T6(3)=18 & T6(4)=12 & T6(5)=27 & T6(6)=6 &
  T7(0)=26 & T7(1)=22 & T7(2)=17 & T7(3)=18 & T7(4)=12 & T7(5)=27 & T7(6)=8 &
  
  A0(0)=0 & A0(1)=0 & A0(2)=0 & A0(3)= 1 & A0(4)= 0 & A0(5)=-1 & A0( 6)=0 &
  A1(0)=0 & A1(1)=0 & A1(2)=0 & A1(3)= 1 & A1(4)= 0 & A1(5)=-1 & A1( 6)=0 &
  A2(0)=0 & A2(1)=0 & A2(2)=1 & A2(3)=-1 & A2(4)= 0 & A2(5)= 1 & A2( 6)=0 &
  A3(0)=0 & A3(1)=0 & A3(2)=1 & A3(3)=-1 & A3(4)= 0 & A3(5)= 0 & A3( 6)=0 &
  A4(0)=0 & A4(1)=1 & A4(2)=0 & A4(3)= 0 & A4(4)=-1 & A4(5)= 0 & A4( 6)=0 &
  A5(0)=0 & A5(1)=1 & A5(2)=0 & A5(3)= 0 & A5(4)=-1 & A5(5)= 0 & A5( 6)=0 &
  A6(0)=0 & A6(1)=0 & A6(2)=0 & A6(3)=-1 & A6(4)= 0 & A6(5)= 0 & A6( 6)=0 &
  A7(0)=0 & A7(1)=1 & A7(2)=0 & A7(3)=-1 & A7(4)= 0 & A7(5)= 0 & A7( 6)=0 &
  
  B0(0)=0 & B0(1)=0 & B0(2)=0 & B0(3)= 1 & B0(4)= 0 & B0(5)=-1 & B0( 6)=0 &
  B1(0)=0 & B1(1)=0 & B1(2)=0 & B1(3)= 1 & B1(4)= 0 & B1(5)=-1 & B1( 6)=0 &
  B2(0)=0 & B2(1)=0 & B2(2)=1 & B2(3)=-1 & B2(4)= 0 & B2(5)= 0 & B2( 6)=0 &
  B3(0)=0 & B3(1)=0 & B3(2)=1 & B3(3)=-1 & B3(4)= 0 & B3(5)= 1 & B3( 6)=0 &
  B4(0)=0 & B4(1)=1 & B4(2)=0 & B4(3)=-0 & B4(4)=-1 & B4(5)= 0 & B4( 6)=0 &
  B5(0)=0 & B5(1)=1 & B5(2)=0 & B5(3)= 0 & B5(4)=-1 & B5(5)= 0 & B5( 6)=0 &
  B6(0)=0 & B6(1)=1 & B6(2)=0 & B6(3)=-1 & B6(4)= 0 & B6(5)= 0 & B6( 6)=0 &
  B7(0)=0 & B7(1)=0 & B7(2)=0 & B7(3)=-1 & B7(4)= 0 & B7(5)= 0 & B7( 6)=0 &
  
  LA=7 & LB=7
  
VARIABLES 
  P0,P1,P2,P3,P4,P5,P6,P7,RA,RB

INVARIANT 
  P0:0..6 & P1:0..6 & P2:0..6 & P3:0..6 & P4:0..6 & P5:0..6 & P6:0..6 & P7:0..6 &
  RA:0..8 & RB:0..8

INITIALISATION 
  P0:=0; P1:=0; P2:=0; P3:=0; P4:=0; P5:=0; P6:=0; P7:=0; RA:=1; RB:=1

OPERATIONS
  move0 = 
    PRE P0<6 & 
      T0(P0+1) /= T1(P1) & 
      T0(P0+1) /= T2(P2 )& 
      T0(P0+1) /= T3(P3 )& 
      T0(P0+1) /= T4(P4) & 
      T0(P0+1) /= T5(P5) & 
      T0(P0+1) /= T6(P6) & 
      T0(P0+1) /= T7(P7) &
      RA + A0(P0+1) <= LA &
      RB + B0(P0+1) <= LB
    THEN 
      P0 := P0+1;
      RA := RA + A0(P0);
      RB := RB + B0(P0)
    END ;

  move1 = 
    PRE P1<6 & 
      T1(P1+1) /= T0(P0) & 
      T1(P1+1) /= T2(P2) &
      T1(P1+1) /= T3(P3) & 
      T1(P1+1) /= T4(P4) & 
      T1(P1+1) /= T5(P5) & 
      T1(P1+1) /= T6(P6) & 
      T1(P1+1) /= T7(P7) &
      RA + A1(P1+1) <= LA &
      RB + B1(P1+1) <= LB
    THEN 
      P1 := P1+1;
      RA := RA + A1(P1);
      RB := RB + B1(P1)
    END ;
               
  move2 = 
    PRE P2<6 & 
      T2(P2+1) /= T0(P0) & 
      T2(P2+1) /= T1(P1) &
      T2(P2+1) /= T3(P3) & 
      T2(P2+1) /= T4(P4) & 
      T2(P2+1) /= T5(P5) & 
      T2(P2+1) /= T6(P6) & 
      T2(P2+1) /= T7(P7) &
      RA + A2(P2+1) <= LA &
      RB + B2(P2+1) <= LB
    THEN 
      P2 := P2+1;
      RA := RA + A2(P2);
      RB := RB + B2(P2)
    END ;
               
  move3 = 
    PRE P3<6 & 
      T3(P3+1) /= T0(P0) & 
      T3(P3+1) /= T1(P1) &
      T3(P3+1) /= T2(P2) & 
      T3(P3+1) /= T4(P4) & 
      T3(P3+1) /= T5(P5) & 
      T3(P3+1) /= T6(P6) & 
      T3(P3+1) /= T7(P7) &
      RA + A3(P3+1) <= LA &
      RB + B3(P3+1) <= LB
    THEN 
      P3 := P3+1;
      RA := RA + A3(P3);
      RB := RB + B3(P3)
    END ;
    
  move4 = 
    PRE P4<6 & 
      T4(P4+1) /= T0(P0) & 
      T4(P4+1) /= T1(P1) & 
      T4(P4+1) /= T2(P2) & 
      T4(P4+1) /= T3(P3) & 
      T4(P4+1) /= T5(P5) & 
      T4(P4+1) /= T6(P6) & 
      T4(P4+1) /= T7(P7) &
      RA + A4(P4+1) <= LA &
      RB + B4(P4+1) <= LB
    THEN 
      P4 := P4+1;
      RA := RA + A4(P4);
      RB := RB + B4(P4)
    END ;

  move5 = 
    PRE P5<6 & 
      T5(P5+1) /= T0(P0) & 
      T5(P5+1) /= T1(P1) &  
      T5(P5+1) /= T2(P2) &
      T5(P5+1) /= T3(P3) & 
      T5(P5+1) /= T4(P4) & 
      T5(P5+1) /= T6(P6) & 
      T5(P5+1) /= T7(P7) &
      RA + A5(P5+1) <= LA &
      RB + B5(P5+1) <= LB
    THEN 
      P5 := P5+1;
      RA := RA + A5(P5);
      RB := RB + B5(P5)
    END ;
    
  move6 = 
    PRE P6<6 & 
      T6(P6+1) /= T0(P0) & 
      T6(P6+1) /= T1(P1) & 
      T6(P6+1) /= T2(P2) & 
      T6(P6+1) /= T3(P3) & 
      T6(P6+1) /= T4(P4) & 
      T6(P6+1) /= T5(P5) & 
      T6(P6+1) /= T7(P7) &
      RA + A6(P6+1) <= LA &
      RB + B6(P6+1) <= LB
    THEN 
      P6 := P6+1;
      RA := RA + A6(P6);
      RB := RB + B6(P6)
    END ;
    
  move7 = 
    PRE P7<6 & 
      T7(P7+1) /= T0(P0) & 
      T7(P7+1) /= T1(P1) & 
      T7(P7+1) /= T2(P2) & 
      T7(P7+1) /= T3(P3) & 
      T7(P7+1) /= T4(P4) & 
      T7(P7+1) /= T5(P5) & 
      T7(P7+1) /= T6(P6) &
      RA + A7(P7+1) <= LA &
      RB + B7(P7+1) <= LB
    THEN 
      P7 := P7+1;
      RA := RA + A7(P7);
      RB := RB + B7(P7)
    END ;
	
  arrived = 
    PRE 
       P0=6 & P1=6 & P2=6 & P3=6 & P4=6 & P5=6 & P6=6 & P7=6 
    THEN 
       skip 
    END
END
 
//--------------------
//  SEARCHING DEADLOCKS:  1_636_547 states,   7_134_235 trans.  TIME  32 min   VMEM  3 GB
//-------------------

\end{verbatim}
}

\section{NuSMV/nuXmv}

{\scriptsize
\begin{verbatim}
MODULE main
 
  -------  train missions ------------------
DEFINE 
T0 := [ 1, 9,10,13,15,20,23]; -- G1
T1 := [ 3, 9,10,13,15,20,24]; -- R1
T2 := [ 5,27,11,13,16,20,25]; -- Y1
T3 := [ 7,27,11,13,16,20,26]; -- B1
T4 := [23,22,17,18,11, 9, 2]; -- G2
T5 := [24,22,17,18,11, 9, 4]; -- R2
T6 := [25,22,17,18,12,27, 6]; -- Y2
T7 := [26,22,17,18,12,27, 8]; -- B2

  ------ region A: train constraints ------
A0 := [ 0, 0, 0, 1, 0,-1, 0]; -- G1
A1 := [ 0, 0, 0, 1, 0,-1, 0]; -- R1
A2 := [ 0, 0, 1,-1, 0, 1, 0]; -- Y1
A3 := [ 0, 0, 1,-1, 0, 0, 0]; -- B1
A4 := [ 0, 1, 0, 0,-1, 0, 0]; -- G2
A5 := [ 0, 1, 0, 0,-1, 0, 0]; -- R2
A6 := [ 0, 0, 0,-1, 0, 0, 0]; -- Y2
A7 := [ 0, 1, 0,-1, 0, 0, 0]; -- B2 
 ------------------------------------------

  ------- region B: train constraints ------
B0 := [ 0, 0, 0, 1, 0,-1, 0]; -- G1
B1 := [ 0, 0, 0, 1, 0,-1, 0]; -- R1
B2 := [ 0, 0, 1,-1, 0, 0, 0]; -- Y1
B3 := [ 0, 0, 1,-1, 0, 1, 0]; -- B1
B4 := [ 0, 1, 0, 0,-1, 0, 0]; -- G2
B5 := [ 0, 1, 0, 0,-1, 0, 0]; -- R2
B6 := [ 0, 1, 0,-1, 0, 0, 0]; -- Y2
B7 := [ 0, 0, 0,-1, 0, 0, 0]; -- B2
  ------------------------------------------
 
LA := 7;
LB := 7;
  
IVAR
  -- (unfair) selector of the train transition
  RUNNING: {0,1,2,3,4,5,6,7};

VAR
  -- vector of train progesses in the execution of their missions
  P0: 0..6;
  P1: 0..6;
  P2: 0..6;
  P3: 0..6;
  P4: 0..6;
  P5: 0..6;
  P6: 0..6;
  P7: 0..6;
  --
  --
  -- the occupation status for regions A and B
  RA:  0..8; 
  RB:  0..8;

ASSIGN
  -- the initial vector of train progesses
  init(P0) := 0;
  init(P1) := 0;
  init(P2) := 0;
  init(P3) := 0;
  init(P4) := 0;
  init(P5) := 0;
  init(P6) := 0;
  init(P7) := 0;
  --
  -- the initial occupation status for regions A and B
  init(RA) := 1;
  init(RB) := 1;
  
TRANS 
  -- progression rules for the evolving train 0
 
      RUNNING =0 &
        -- the current train has not yet completed its mission
        P0 < 6 &
        --
        -- the next place is not occupied by other trains
        T0[P0+1] !=  T1[P1] &
        T0[P0+1] !=  T2[P2] &
        T0[P0+1] !=  T3[P3] &
        T0[P0+1] !=  T4[P4] &
        T0[P0+1] !=  T5[P5] &
        T0[P0+1] !=  T6[P6] &
        T0[P0+1] !=  T7[P7] &
        -- 
        -- the progression step of id satisfies all contraints
        RA + A0[P0+1] <= LA &
        RB + B0[P0+1] <= LB 
      ?
        next(P0) in (P0+1) &
        next(P1) in P1 &
        next(P2) in P2 &
        next(P3) in P3 &
        next(P4) in P4 &
        next(P5) in P5 &
        next(P6) in P6 &
        next(P7) in P7 &
        next(RA) in (RA + A0[P0+1]) &
        next(RB) in (RB + B0[P0+1]) 
      : 
      RUNNING = 1 &
        P1 < 6 &
        T1[P1+1] !=  T0[P0] &
        T1[P1+1] !=  T2[P2] &
        T1[P1+1] !=  T3[P3] &
        T1[P1+1] !=  T4[P4] &
        T1[P1+1] !=  T5[P5] &
        T1[P1+1] !=  T6[P6] &
        T1[P1+1] !=  T7[P7] &
        RA + A1[P1+1] <= LA &
        RB + B1[P1+1] <= LB 
      ? 
        next(P0) in P0 &
        next(P1) in (P1+1) &
        next(P2) in P2 &
        next(P3) in P3 &
        next(P4) in P4 &
        next(P5) in P5 &
        next(P6) in P6 &
        next(P7) in P7 &
        next(RA) in (RA + A1[P1+1]) &
        next(RB) in (RB + B1[P1+1])    
     : 
      RUNNING =2 &
        P2 < 6 &
        T2[P2+1] !=  T0[P0] &
        T2[P2+1] !=  T1[P1] &
        T2[P2+1] !=  T3[P3] &
        T2[P2+1] !=  T4[P4] &
        T2[P2+1] !=  T5[P5] &
        T2[P2+1] !=  T6[P6] &
        T2[P2+1] !=  T7[P7] &
        RA + A2[P2+1] <= LA &
        RB + B2[P2+1] <= LB
      ?
        next(P0) in P0 &
        next(P1) in P1 &
        next(P2) in (P2+1) &
        next(P3) in P3 &
        next(P4) in P4 &
        next(P5) in P5 &
        next(P6) in P6 &
        next(P7) in P7 &
        next(RA) in (RA + A2[P2+1]) &
        next(RB) in (RB + B2[P2+1])    
    :
     RUNNING =3 &
        P3 < 6 &
        T3[P3+1] !=  T0[P0] &
        T3[P3+1] !=  T1[P1] &
        T3[P3+1] !=  T2[P2] &
        T3[P3+1] !=  T3[P3] &
        T3[P3+1] !=  T4[P4] &
        T3[P3+1] !=  T5[P5] &
        T3[P3+1] !=  T6[P6] &
        T3[P3+1] !=  T7[P7] &
        RA + A3[P3+1] <= LA &
        RB + B3[P3+1] <= LB 
    ?
        next(P0) in P0 &
        next(P1) in P1 &
        next(P2) in P2 &
        next(P3) in (P3+1) &
        next(P4) in P4 &
        next(P5) in P5 &
        next(P6) in P6 &
        next(P7) in P7 &
        next(RA) in (RA + A3[P3+1]) &
        next(RB) in (RB + B3[P3+1]) 
   
    :
     RUNNING =4 &
        P4 < 6 &
        T4[P4+1] !=  T0[P0] &
        T4[P4+1] !=  T1[P1] &
        T4[P4+1] !=  T2[P2] &
        T4[P4+1] !=  T3[P3] &
        T4[P4+1] !=  T4[P4] &
        T4[P4+1] !=  T5[P5] &
        T4[P4+1] !=  T6[P6] &
        T4[P4+1] !=  T7[P7] &
        RA + A4[P4+1] <= LA &
        RB + B4[P4+1] <= LB
    ?
        next(P0) in P0 &
        next(P1) in P1 &
        next(P2) in P2 &
        next(P3) in P3 &
        next(P4) in (P4+1) &
        next(P5) in P5 &
        next(P6) in P6 &
        next(P7) in P7 &
        next(RA) in (RA + A4[P4+1]) &
        next(RB) in (RB + B4[P4+1]) 
    :
     RUNNING =5 &
        P5 < 6 &
        T5[P5+1] !=  T0[P0] &
        T5[P5+1] !=  T1[P1] &
        T5[P5+1] !=  T2[P2] &
        T5[P5+1] !=  T3[P3] &
        T5[P5+1] !=  T4[P4] &
        T5[P5+1] !=  T6[P6] &
        T5[P5+1] !=  T7[P7] &
        RA + A5[P5+1] <= LA &
        RB + B5[P5+1] <= LB 
    ?
        next(P0) in P0 &
        next(P1) in P1 &
        next(P2) in P2 &
        next(P3) in P3 &
        next(P4) in P4 &
        next(P5) in (P5+1) &
        next(P6) in P6 &
        next(P7) in P7 &
        next(RA) in (RA + A5[P5+1]) &
        next(RB) in (RB + B5[P5+1]) 
    :
      RUNNING = 6 &
        P6 < 6 &
        T6[P6+1] !=  T0[P0] &
        T6[P6+1] !=  T1[P1] &
        T6[P6+1] !=  T2[P2] &
        T6[P6+1] !=  T3[P3] &
        T6[P6+1] !=  T4[P4] &
        T6[P6+1] !=  T5[P5] &
        T6[P6+1] !=  T6[P6] &
        T6[P6+1] !=  T7[P7] &
        RA + A6[P6+1] <= LA &
        RB + B6[P6+1] <= LB
    ?
        next(P0) in P0 &
        next(P1) in P1 &
        next(P2) in P2 &
        next(P3) in P3 &
        next(P4) in P4 &
        next(P5) in P5 &
        next(P6) in (P6+1) &
        next(P7) in P7 &
        next(RA) in (RA + A6[P6+1]) &
        next(RB) in (RB + B6[P6+1])  
    :
     RUNNING =7 &
        -- the current train has not yet completed its mission
        P7 < 6 &
        --
        -- the next place is not occupied by other trains
        T7[P7+1] !=  T0[P0] &
        T7[P7+1] !=  T1[P1] &
        T7[P7+1] !=  T2[P2] &
        T7[P7+1] !=  T3[P3] &
        T7[P7+1] !=  T4[P4] &
        T7[P7+1] !=  T5[P5] &
        T7[P7+1] !=  T6[P6] &
        T7[P7+1] !=  T7[P7] &
        -- 
        -- the progression step of id satisfies all contraints
        RA + A7[P7+1] <= LA &
        RB + B7[P7+1] <= LB 
    ?
        next(P0) in P0 &
        next(P1) in P1 &
        next(P2) in P2 &
        next(P3) in P3 &
        next(P4) in P4 &
        next(P5) in P5 &
        next(P6) in P6 &
        next(P7) in (P7+1) &
        next(RA) in (RA + A7[P7+1]) &
        next(RB) in (RB + B7[P7+1])
   :  
        next(P0) in P0 &
        next(P1) in P1 &
        next(P2) in P2 &
        next(P3) in P3 &
        next(P4) in P4 &
        next(P5) in P5 &
        next(P6) in P6 &
        next(P7) in P7 &
        next(RA) in RA &
        next(RB) in RB

-- FAIRNESS  RUNNING = 0;
-- FAIRNESS  RUNNING = 1;
-- FAIRNESS  RUNNING = 2;
-- FAIRNESS  RUNNING = 3;
-- FAIRNESS  RUNNING = 4;
-- FAIRNESS  RUNNING = 5;
-- FAIRNESS  RUNNING = 6;
-- FAIRNESS  RUNNING = 7;
 
-- CTLSPEC
--   AF ((P0=6) & (P1=6) & (P2=6) & (P3=6) &  (P4=6) & (P5=6) & (P6=6) & (P7=6))
   
-- LTLSPEC
--   F ((P0=6) & (P1=6) & (P2=6) & (P3=6) &  (P4=6) & (P5=6) & (P6=6) & (P7=6))

CTLSPEC 
  AG EF ((P0=6) & (P1=6) & (P2=6) & (P3=6) &  (P4=6) & (P5=6) & (P6=6) & (P7=6))
  
-------------------------------- end main -------------------------------------------
-------------------------------------------------------------------------------------

----------------------------------------
--    Batch Verification:
----------------------------------------
-- time nusmv -r -v 1 smv_oneway8-SM.smv
--    FAIRNESS RUNNING = 1; ... RUNNING = 7;
--    LTLSPEC    F ((0=6) & ... & P7=6))
-- > 
-- > reachable states: 1.63654e+06 (2^20.6422) out of 4.66949e+08 (2^28.7987)
-- > Successful termination
-- > real	0m43.609s
-- > user	0m43.431s
----------------------------------------
-- time nusmv -r -v 1 smv_oneway8-SM.smv
--    FAIRNESS RUNNING = 1; ... RUNNING = 7;
--    CTLSPEC   AF ((0=6) & ... & P7=6))
-- > 
-- > reachable states: 1.63654e+06 (2^20.6422) out of 4.66949e+08 (2^28.7987)
-- > Successful termination
-- > real	0m39.211s
-- > user	0m39.015s
----------------------------------------
-- time nusmv -r -v 1 smv_oneway8-SM.smv
--    CTLSPEC   AG EF ((0=6) & ... & P7=6))
-- > 
-- > reachable states: 1.63654e+06 (2^20.6422) out of 4.66949e+08 (2^28.7987)
-- > Successful termination
-- > real	0m2.807s
-- > user	0m2.771s
-- > USED MEMORY  74 MB

----------------------------------------
-- nusmv -v 2 -ctt -r -is smv_oneway8-SM.smv
--    -ctt checks totatlity of transition relation function
--    -r  prints actual number of reachable states 
--    -v 1   verbose (1..4)
--    -is   ignore SPEC properties
--    -AG   used ad hoc algorithm for AG-only properties
-- nusmv -v 1 -bmc -bmc_length 100 cyclic8-smv.txt 
------------------------------------------
--    Interactive Verification:
-- ./NuSMV -int
-- read_model -i smv_oneway8-SM.smv
-- flatten_hierarchy
-- encode_variables
-- build_model
-- check_ctlspec -p "AF (P0 = 0)"
--------------- other commands --------
--check_ctlspec [-h] [-m | -o output-file] [-n number | -p
-- "ctl-expr  [IN context]" | -P "name"]
--go
--pick_state -i
--simulate -i
---------------------------------------
\end{verbatim}
}

\section{SPIN}

{\scriptsize
\begin{verbatim}

/*  TRAIN MISSION DATA */
byte T0[14];
byte T1[14];
byte T2[14];
byte T3[14];
byte T4[14];
byte T5[14];
byte T6[14];
byte T7[14];

/* TRAIN PROGRESS DATA */
byte P0,P1,P2,P3,P4,P5,P6,P7;

/*  CONSTRAINTs DATA  FOR REGIONS A,B  */
byte RA; // occupancy of region A
byte RB; // occupancy of region B
byte LA; // limit of region A
byte LB; // limit if region B
short A0[14]; // Constraints of Train 0 for Region A
short A1[14]; // Constraints of Train 1 for Region A
short A2[14]; // Constraints of Train 2 for Region A
short A3[14]; //          ...
short A4[14];
short A5[14];
short A6[14];
short A7[14];
short B0[14]; // Constraints of Train 0 for Region B
short B1[14]; // Constraints of Train 1 for Region B
short B2[14]; // Constraints of Train 2 for Region B
short B3[14]; //          ...
short B4[14];
short B5[14];
short B6[14];
short B7[14];

/*  INITIALIZATIONS */
init {
  atomic {

//--------------------------------------------------------------
// T0 := [ 1, 9,10,13,15,20,23,22]; -- G1
// T1 := [ 3, 9,10,13,15,20,24,22]; -- R1
// T2 := [ 5,27,11,13,16,20,25,22]; -- Y1
// T3 := [ 7,27,11,13,16,20,26,22]; -- B1
// T4 := [23,22,17,18,11, 9, 2, 1]; -- G2
// T5 := [24,22,17,18,11, 9, 4, 3]; -- R2
// T6 := [25,22,17,18,12,27, 6, 5]; -- Y2
// T7 := [26,22,17,18,12,27, 8, 7]; -- B2
//--------------------------------------------------------------
  T0[0]= 1; T0[1]= 9; T0[2]=10; T0[3]=13; T0[4]=15; T0[5]=20; T0[6]=23;
  T1[0]= 3; T1[1]=9;  T1[2]=10; T1[3]=13; T1[4]=15; T1[5]=20; T1[6]=24;
  T2[0]= 5; T2[1]=27; T2[2]=11; T2[3]=13; T2[4]=16; T2[5]=20; T2[6]=25;
  T3[0]= 7; T3[1]=27; T3[2]=11; T3[3]=13; T3[4]=16; T3[5]=20; T3[6]=26;
  T4[0]=23; T4[1]=22; T4[2]=17; T4[3]=18; T4[4]=11; T4[5]= 9; T4[6]= 2;
  T5[0]=24; T5[1]=22; T5[2]=17; T5[3]=18; T5[4]=11; T5[5]= 9; T5[6]= 4;
  T6[0]=25; T6[1]=22; T6[2]=17; T6[3]=18; T6[4]=12; T6[5]=27; T6[6]= 6;
  T7[0]=26; T7[1]=22; T7[2]=17; T7[3]=18; T7[4]=12; T7[5]=27; T7[6]= 8;

//  
// ------ initial train positions   --------
//  Pi=0 as default value. no need of explicit initialization

// ------ region A: train constraints ------
   A0[3] = 1; A0[5] = -1; A0[ 7]=  1; A0[10] = -1;
   A1[3] = 1; A1[5] = -1; A1[ 7]=  1; A1[10] = -1;
   A2[2] = 1; A2[3] = -1; A2[ 5]=  1; A2[ 9] = -1;
   A3[2] = 1; A3[3] = -1; A3[ 7]=  1; A3[ 9] = -1;
   A4[1] = 1; A4[4] = -1; A4[10]=  1; A4[12] = -1;
   A5[1] = 1; A5[4] = -1; A5[10]=  1; A5[12] = -1;
   A6[3] =-1; A6[ 9] = 1; A6[10]= -1; A6[12] =  1;
   A7[1] = 1; A7[3] = -1; A7[ 9]=  1; A7[10] = -1;
   
// ------- region B: train constraints ------
   B0[3] =  1; B0[5] = -1; B0[ 7] =  1; B0[10] = -1;
   B1[3] =  1; B1[5] = -1; B1[ 7] =  1; B1[10] = -1;
   B2[2] =  1; B2[3] = -1; B2[ 7] =  1; B2[ 9] = -1;
   B3[2] =  1; B3[3] = -1; B3[ 5] =  1; B3[ 9] = -1;
   B4[1] =  1; B4[4] = -1; B4[10] =  1; B4[12] = -1;
   B5[1] =  1; B5[4] = -1; B5[10] =  1; B5[12] = -1;
   B6[1] =  1; B6[3] = -1; B6[ 9] =  1; B6[10] = -1;
   B7[3] = -1; B7[9] =  1; B7[10] = -1; B7[12] =  1;

   RA = 1;
   RB = 1;
   LA =7;
   LB =7;
  }
   do
    :: atomic {
       (P0 < 6 &&
        T0[P0+1] != T1[P1] && // next place of train0 not occupied by train1
        T0[P0+1] != T2[P2] && // next place of train0 not occupied by train2
        T0[P0+1] != T3[P3] &&
        T0[P0+1] != T4[P4] &&
        T0[P0+1] != T5[P5] &&
        T0[P0+1] != T6[P6] &&
        T0[P0+1] != T7[P7] && // next place of train0 not occupied by train7
        (RA + A0[P0+1]) <= LA && // progress of train0 does not saturate RA 
        (RB + B0[P0+1]) <= LB    // progress of train0 does not saturate RB 
       ) ->  
         P0 = (P0+1); 
         RA = RA + A0[P0];   // update occupancy of RA according to the step
         RB = RB + B0[P0];   // update occupancy of RB according to the step
       };  
    :: atomic {
       (P1 < 6 &&
        T1[P1+1] != T0[P0] &&
        T1[P1+1] != T2[P2] &&
        T1[P1+1] != T3[P3] &&
        T1[P1+1] != T4[P4] &&
        T1[P1+1] != T5[P5] &&
        T1[P1+1] != T6[P6] &&
        T1[P1+1] != T7[P7] &&
        (RA + A1[P1+1]) <= LA &&
        (RB + B1[P1+1]) <= LB    // progress of train0 does not saturate RD 
       ) ->  
         P1 = (P1+1); 
         RA = RA + A1[P1];  
         RB = RB + B1[P1];  
       };
    :: atomic {
       (P2 < 6 &&
        T2[P2+1] != T0[P0] &&
        T2[P2+1] != T1[P1] &&
        T2[P2+1] != T3[P3] &&
        T2[P2+1] != T4[P4] &&
        T2[P2+1] != T5[P5] &&
        T2[P2+1] != T6[P6] &&
        T2[P2+1] != T7[P7] &&
        (RA + A2[P2+1]) <= LA &&
        (RB + B2[P2+1]) <= LB 
       ) ->  
         P2 = (P2+1); 
         RA = RA + A2[P2];   // update occupancy of RA according to the step
         RB = RB + B2[P2]; 
       };
    :: atomic {
       (P3 < 6 &&
        T3[P3+1] != T0[P0] &&
        T3[P3+1] != T1[P1] &&
        T3[P3+1] != T2[P2] &&
        T3[P3+1] != T4[P4] &&
        T3[P3+1] != T5[P5] &&
        T3[P3+1] != T6[P6] &&
        T3[P3+1] != T7[P7] &&
        (RA + A3[P3+1]) <= LA &&
        (RB + B3[P3+1]) <= LB 
       ) ->  
         P3 = (P3+1); 
         RA = RA + A3[P3];   // update occupancy of RA according to the step
         RB = RB + B3[P3]; 
       };
    :: atomic {
       (P4 < 6 &&
        T4[P4+1] != T0[P0] &&
        T4[P4+1] != T1[P1] &&
        T4[P4+1] != T2[P2] &&
        T4[P4+1] != T3[P3] &&
        T4[P4+1] != T5[P5] &&
        T4[P4+1] != T6[P6] &&
        T4[P4+1] != T7[P7] &&
        (RA + A4[P4+1]) <= LA &&
        (RB + B4[P4+1]) <= LB 
       ) ->  
         P4 = (P4+1); 
         RA = RA + A4[P4];   // update occupancy of RA according to the step
         RB = RB + B4[P4];
       };
    :: atomic {
       (P5 < 6 &&
        T5[P5+1] != T0[P0] &&
        T5[P5+1] != T1[P1] &&
        T5[P5+1] != T2[P2] &&
        T5[P5+1] != T3[P3] &&
        T5[P5+1] != T4[P4] &&
        T5[P5+1] != T6[P6] &&
        T5[P5+1] != T7[P7] &&
        (RA + A5[P5+1]) <= LA &&
        (RB + B5[P5+1]) <= LB  
       ) ->  
         P5 = (P5+1);  
         RA = RA + A5[P5];   // update occupancy of RA according to the step
         RB = RB + B5[P5]; 
       };
    :: atomic {
       (P6 < 6 &&
        T6[P6+1] != T0[P0] &&
        T6[P6+1] != T1[P1] &&
        T6[P6+1] != T2[P2] &&
        T6[P6+1] != T3[P3] &&
        T6[P6+1] != T4[P4] &&
        T6[P6+1] != T5[P5] &&
        T6[P6+1] != T7[P7] &&
        (RA + A6[P6+1]) <= LA &&
        (RB + B6[P6+1]) <= LB 
       ) ->  
         P6 = (P6+1); 
         RA = RA + A6[P6];   // update occupancy of RA according to the step
         RB = RB + B6[P6]; 
       };
    :: atomic {
       (P7 < 6 &&
        T7[P7+1] != T0[P0] &&
        T7[P7+1] != T1[P1] &&
        T7[P7+1] != T2[P2] &&
        T7[P7+1] != T3[P3] &&
        T7[P7+1] != T4[P4] &&
        T7[P7+1] != T5[P5] &&
        T7[P7+1] != T6[P6] &&
        (RA + A7[P7+1]) <= LA &&
        (RB + B7[P7+1]) <= LB 
       ) ->  
         P7 = (P7+1); 
         RA = RA + A7[P7];   // update occupancy of RA according to the step
         RB = RB + B7[P7]; 
       };
 :: (P0 == 6) && (P1 == 6) && (P2 == 6) && (P3 == 6) &&
    (P4 == 6) && (P5 == 6) && (P6 == 6) && (P7 == 6) -> skip;
  od;
}

/*  PROPERTIES */
 ltl p1 
 { <> ((P0==6) && (P1==6) && (P2==6) && (P3==6) &&
       (P4==6) && (P5==6) && (P6==6) && (P7==6)) }
       
/*   verfication process
  // DEPTH FIRST 
  spin -a spin_oneway8small.pml
  gcc -O3 -o pan pan.c 
  time pan -a 
  > 
  > Full statespace search for:
  >     never claim         	+ (p1)
  >
  >     1636546 states, stored
  >     7134234 transitions
  >   
  > real	0m13.110s
  > user	0m12.683s
  > sys	0m0.411s
  > USED VIRTUAL MEMORY (pan):  1.02 GB

  // BREADTH FIRST
  spin -a spin_oneway8.pml
  gcc -O3 -DBFS -DBFS_DISK -DVECTORSZ=256000  -o pan pan.c
  gcc -O3 -DBFS -DVECTORSZ=256000  -o pan pan.c 
  time pan -m500000 -v -w33
  >
  > Full statespace search for:
  >	never claim         	+ (p1)
  >    
  >    1636545 states, stored
  >     7134237 transitions
  >
  > real	1m3.582s
  > user	0m31.621s
  > sys	 	0m29.806s

*/

/*   other commands
  spin -t[N] -- follow [Nth] simulation trail, see also -k
  pan -c0  -- counts all errors
  pan -c   -- saves in the trail file the info for 3rd error
  pan -e -c0 -- saves all errors trails each one in file specI.trail
  spin -k specI.trail -c spec.pml -- displays the trail for error I
  pan -r trailfilename   --read and execute trail in file
  pan -rN    --   read and execute N-th error trail
  pan -C  --  read and execute trail - columnated output (can add -v,-n)
  pan -r -PN read and execute trail - restrict trail output to proc N
  pan -     (for help on options)  
  pan -w32 -v  -D (dot format!)
  ------
*/

\end{verbatim}
}

\section{TLA+}

{\scriptsize
\begin{verbatim}
------------------ MODULE oneway ---------------
EXTENDS Integers

VARIABLE 
  P0,P1,P2,P3,P4,P5,P6,P7,RA,RB
  
vars == <<P0,P1,P2,P3,P4,P5,P6,P7,RA,RB>>

TypesOK  == (P0 \in 0..6)

T0 == << 1, 9,10,13,15,20,23>>
T1 == << 3, 9,10,13,15,20,24>>
T2 == << 5,27,11,13,16,20,25>>
T3 == << 7,27,11,13,16,20,26>>
T4 == <<23,22,17,18,11, 9, 2>>
T5 == <<24,22,17,18,11, 9, 4>>
T6 == <<25,22,17,18,12,27, 6>>
T7 == <<26,22,17,18,12,27, 8>>

A0 == << 0, 0, 0, 1, 0,-1, 0>>
A1 == << 0, 0, 0, 1, 0,-1, 0>>
A2 == << 0, 0, 1,-1, 0, 1, 0>>
A3 == << 0, 0, 1,-1, 0, 0, 0>>
A4 == << 0, 1, 0, 0,-1, 0, 0>>
A5 == << 0, 1, 0, 0,-1, 0, 0>>
A6 == << 0, 0, 0,-1, 0, 0, 0>>
A7 == << 0, 1, 0,-1, 0, 0, 0>>

B0 == << 0, 0, 0, 1, 0,-1, 0>>
B1 == << 0, 0, 0, 1, 0,-1, 0>>
B2 == << 0, 0, 1,-1, 0, 0, 0>>
B3 == << 0, 0, 1,-1, 0, 1, 0>>
B4 == << 0, 1, 0, 0,-1, 0, 0>>
B5 == << 0, 1, 0, 0,-1, 0, 0>>
B6 == << 0, 1, 0,-1, 0, 0, 0>>
B7 == << 0, 0, 0,-1, 0, 0, 0>>

LA == 7
LB == 7

LL == 6

Init== 
/\ P0=0 /\ P1=0 /\ P2=0 /\ P3=0 /\ P4=0 /\ P5=0 /\ P6=0 /\ P7=0 
/\ RA=1 /\ RB=1

Move0 ==
   /\ P0 < LL 
   /\ T0[P0+2] /= T1[P1+1]
   /\ T0[P0+2] /= T2[P2+1]
   /\ T0[P0+2] /= T3[P3+1]
   /\ T0[P0+2] /= T4[P4+1]
   /\ T0[P0+2] /= T5[P5+1]
   /\ T0[P0+2] /= T6[P6+1]
   /\ T0[P0+2] /= T7[P7+1]
   /\ RA + A0[P0+2] <= LA
   /\ RB + B0[P0+2] <= LB
   /\ P0' = (P0+1)
   /\ RA' = RA + A0[P0+2]
   /\ RB' = RB + B0[P0+2]
   /\ UNCHANGED <<P1,P2,P3,P4,P5,P6,P7>>


Move1 ==
   /\ P1 < LL 
   /\ T1[P1+2] /= T0[P0+1]
   /\ T1[P1+2] /= T2[P2+1]
   /\ T1[P1+2] /= T3[P3+1]
   /\ T1[P1+2] /= T4[P4+1]
   /\ T1[P1+2] /= T5[P5+1]
   /\ T1[P1+2] /= T6[P6+1]
   /\ T1[P1+2] /= T7[P7+1]
   /\ RA + A1[P1+2] <= LA
   /\ RB + B1[P1+2] <= LB
   /\ P1' = (P1+1)
   /\ RA' = RA + A1[P1+2]
   /\ RB' = RB + B1[P1+2]
   /\ UNCHANGED <<P0,P2,P3,P4,P5,P6,P7>>
   

Move2 ==
   /\ P2 < LL 
   /\ T2[P2+2] /= T0[P0+1]
   /\ T2[P2+2] /= T1[P1+1]
   /\ T2[P2+2] /= T3[P3+1]
   /\ T2[P2+2] /= T4[P4+1]
   /\ T2[P2+2] /= T5[P5+1]
   /\ T2[P2+2] /= T6[P6+1]
   /\ T2[P2+2] /= T7[P7+1]
   /\ RA + A2[P2+2] <= LA
   /\ RB + B2[P2+2] <= LB
   /\ P2' = (P2+1)
   /\ RA' = RA + A2[P2+2]
   /\ RB' = RB + B2[P2+2]
   /\ UNCHANGED <<P0,P1,P3,P4,P5,P6,P7>>
   

Move3 ==
   /\ P3 < LL 
   /\ T3[P3+2] /= T0[P0+1]
   /\ T3[P3+2] /= T1[P1+1]
   /\ T3[P3+2] /= T2[P2+1]
   /\ T3[P3+2] /= T4[P4+1]
   /\ T3[P3+2] /= T5[P5+1]
   /\ T3[P3+2] /= T6[P6+1]
   /\ T3[P3+2] /= T7[P7+1]
   /\ RA + A3[P3+2] <= LA
   /\ RB + B3[P3+2] <= LB
   /\ P3' = (P3+1)
   /\ RA' = RA + A3[P3+2]
   /\ RB' = RB + B3[P3+2]
   /\ UNCHANGED <<P0,P1,P2,P4,P5,P6,P7>>
     
Move4 ==
   /\ P4 < LL 
   /\ T4[P4+2] /= T0[P0+1]
   /\ T4[P4+2] /= T1[P1+1]
   /\ T4[P4+2] /= T2[P2+1]
   /\ T4[P4+2] /= T3[P3+1]
   /\ T4[P4+2] /= T5[P5+1]
   /\ T4[P4+2] /= T6[P6+1]
   /\ T4[P4+2] /= T7[P7+1]
   /\ RA + A4[P4+2] <= LA
   /\ RB + B4[P4+2] <= LB
   /\ P4' = (P4+1)
   /\ RA' = RA + A4[P4+2]
   /\ RB' = RB + B4[P4+2]
   /\ UNCHANGED <<P0,P1,P2,P3,P5,P6,P7>>
   

Move5 ==
   /\ P5 < LL 
   /\ T5[P5+2] /= T0[P0+1]
   /\ T5[P5+2] /= T1[P1+1]
   /\ T5[P5+2] /= T2[P2+1]
   /\ T5[P5+2] /= T3[P3+1]
   /\ T5[P5+2] /= T4[P4+1]
   /\ T5[P5+2] /= T6[P6+1]
   /\ T5[P5+2] /= T7[P7+1]
   /\ RA + A5[P5+2] <= LA
   /\ RB + B5[P5+2] <= LB
   /\ P5' = (P5+1)
   /\ RA' = RA + A5[P5+2]
   /\ RB' = RB + B5[P5+2]
   /\ UNCHANGED <<P0,P1,P2,P3,P4,P6,P7>>
   
Move6 ==
   /\ P6 < LL 
   /\ T6[P6+2] /= T0[P0+1]
   /\ T6[P6+2] /= T1[P1+1]
   /\ T6[P6+2] /= T2[P2+1]
   /\ T6[P6+2] /= T3[P3+1]
   /\ T6[P6+2] /= T4[P4+1]
   /\ T6[P6+2] /= T5[P5+1]
   /\ T6[P6+2] /= T7[P7+1]
   /\ RA + A6[P6+2] <= LA
   /\ RB + B6[P6+2] <= LB
   /\ P6' = (P6+1)
   /\ RA' = RA + A6[P6+2]
   /\ RB' = RB + B6[P6+2]
   /\ UNCHANGED <<P0,P1,P2,P3,P4,P5,P7>>
   
   
Move7 ==
   /\ P7 < LL 
   /\ T7[P7+2] /= T0[P0+1]
   /\ T7[P7+2] /= T1[P1+1]
   /\ T7[P7+2] /= T2[P2+1]
   /\ T7[P7+2] /= T3[P3+1]
   /\ T7[P7+2] /= T4[P4+1]
   /\ T7[P7+2] /= T5[P5+1]
   /\ T7[P7+2] /= T6[P6+1]
   /\ RA + A7[P7+2] <= LA
   /\ RB + B7[P7+2] <= LB
   /\ P7' = (P7+1)
   /\ RA' = RA + A7[P7+2]
   /\ RB' = RB + B7[P7+2]
   /\ UNCHANGED <<P0,P1,P2,P3,P4,P5,P6>>
   
Next== 
  \/  Move0 \/  Move1 \/  Move2 \/  Move3
  \/  Move4 \/  Move5 \/  Move6 \/  Move7
    
Arrived == 
  /\ (P0=LL) /\ (P1=LL) /\ (P2=LL) /\ (P3=LL)
  /\ (P4=LL) /\ (P5=LL) /\ (P6=LL) /\ (P7=LL)
 
Property == <>Arrived
  
Spec == Init /\ [][Next]_vars 
SFairSpec == Init /\ [][Next]_vars /\ SF_vars (Next)    (*for LTL verification*)

(**************************************************)
(* Property: <>Arrived,  Behavior Spec: SFairSpec *)
(* States: 1636545, Result: TRUE, Time 3m17s      *)
(**************************************************)

(* Model Overview:  setting Temporal formula == "Spec" *)
(* Deadlock Found:  trace for  P0=6 & P4=6 *)
(* PROPERTIES: <>Arrivedis FALSE, because of implicit stuttering*)

(* Model Overview: setting Temporal formula == "SFairSpec" *)
(* Deadlock Found: trace for P0=6 & P4=6 (stuttering does not avoids deadlocks)*)
(* PROPERTIES: <>Arrived     is TRUE, stuttering ignored *)
===============================================

\end{verbatim}
}

\section{UMC}

{\scriptsize
\begin{verbatim}
Class REGION2 is
Vars:

---------------------------------------------------------------
T0: int[] := [ 1, 9,10,13,15,20,23]; -- G1
T1: int[] := [ 3, 9,10,13,15,20,24]; -- R1
T2: int[] := [ 5,27,11,13,16,20,25]; -- Y1
T3: int[] := [ 7,27,11,13,16,20,26]; -- B1
T4: int[] := [23,22,17,18,11, 9, 2]; -- G2
T5: int[] := [24,22,17,18,11, 9, 4]; -- R2
T6: int[] := [25,22,17,18,12,27, 6]; -- Y2
T7: int[] := [26,22,17,18,12,27, 8]; -- B2
----------------------------------------------------------------
P0: int :=0;
P1: int :=0;
P2: int :=0;
P3: int :=0;
P4: int :=0;
P5: int :=0;
P6: int :=0;
P7: int :=0;
----------------------------------------------------------------

------ region A: train constraints ------
A0: int[] := [ 0, 0, 0, 1, 0,-1, 0]; -- G1
A1: int[] := [ 0, 0, 0, 1, 0,-1, 0]; -- R1
A2: int[] := [ 0, 0, 1,-1, 0, 1, 0]; -- Y1
A3: int[] := [ 0, 0, 1,-1, 0, 0, 0]; -- B1
A4: int[] := [ 0, 1, 0, 0,-1, 0, 0]; -- G2
A5: int[] := [ 0, 1, 0, 0,-1, 0, 0]; -- R2
A6: int[] := [ 0, 0, 0,-1, 0, 0, 0]; -- Y2
A7: int[] := [ 0, 1, 0,-1, 0, 0, 0]; -- B2
------------------------------------------

------- region B: train constraints ------
B0: int[] := [ 0, 0, 0, 1, 0,-1, 0]; -- G1
B1: int[] := [ 0, 0, 0, 1, 0,-1, 0]; -- R1
B2: int[] := [ 0, 0, 1,-1, 0, 0, 0]; -- Y1
B3: int[] := [ 0, 0, 1,-1, 0, 1, 0]; -- B1
B4: int[] := [ 0, 1, 0, 0,-1, 0, 0]; -- G2
B5: int[] := [ 0, 1, 0, 0,-1, 0, 0]; -- R2
B6: int[] := [ 0, 1, 0,-1, 0, 0, 0]; -- Y2
B7: int[] := [ 0, 0, 0,-1, 0, 0, 0]; -- B2
------------------------------------------


-------------------------------------------------------------
RA: int :=1;  -- initial value for region RA
RB: int :=1;  -- initial value for region RB
LA: int :=7;  -- limit value for region RA
LB: int :=7;  -- limit value for region RB
-------------------------------------------------------------------

State Top =s1

Behavior:

------------------------- train 0 -----------------------------
s1 -> s1
 {- [P0 < 6 &
     T0[P0+1] != T1[P1] &
     T0[P0+1] != T2[P2] &
     T0[P0+1] != T3[P3] &
     T0[P0+1] != T4[P4] &
     T0[P0+1] != T5[P5] &
     T0[P0+1] != T6[P6] &
     T0[P0+1] != T7[P7] &
     RA + A0[P0+1] <= LA &
     RB + B0[P0+1] <= LB] /
  P0 := P0 +1;
  RA = RA + A0[P0];
  RB = RB + B0[P0];
 }

------------------------- train 1 -----------------------------
s1 -> s1 
 {- [P1 < 6 &
     T1[P1+1] != T0[P0] &
     T1[P1+1] != T2[P2] &
     T1[P1+1] != T3[P3] &
     T1[P1+1] != T4[P4] &
     T1[P1+1] != T5[P5] &
     T1[P1+1] != T6[P6] &
     T1[P1+1] != T7[P7] &
     RA + A1[P1+1] <= LA &
     RB + B1[P1+1] <= LB ] /
  P1 := P1 +1;
  RA = RA + A1[P1];
  RB = RB + B1[P1];
 }
     
------------------------- train 2 -----------------------------
s1 -> s1 
 {- [P2 < 6 &
     T2[P2+1] != T0[P0] &
     T2[P2+1] != T1[P1] &
     T2[P2+1] != T3[P3] &
     T2[P2+1] != T4[P4] &
     T2[P2+1] != T5[P5] &
     T2[P2+1] != T6[P6] &
     T2[P2+1] != T7[P7] &
     RA + A2[P2+1] <= LA &
     RB + B2[P2+1] <= LB ] /
  P2 := P2 +1;
  RA = RA + A2[P2];
  RB = RB + B2[P2];
 }

------------------------- train 3 -----------------------------
s1 -> s1 
 {- [P3 < 6 &
     T3[P3+1] != T0[P0] &
     T3[P3+1] != T1[P1] &
     T3[P3+1] != T2[P2] &
     T3[P3+1] != T4[P4] &
     T3[P3+1] != T5[P5] &
     T3[P3+1] != T6[P6] &
     T3[P3+1] != T7[P7] &
     RA + A3[P3+1] <= LA &
     RB + B3[P3+1] <= LB ] /
  P3 := P3 +1;
  RA = RA + A3[P3];
  RB = RB + B3[P3];
 }

------------------------- train 4 -----------------------------
s1 -> s1 
 {- [P4 < 6 &
     T4[P4+1] != T0[P0] &
     T4[P4+1] != T1[P1] &
     T4[P4+1] != T2[P2] &
     T4[P4+1] != T3[P3] &
     T4[P4+1] != T5[P5] &
     T4[P4+1] != T6[P6] &
     T4[P4+1] != T7[P7] &
     RA + A4[P4+1] <= LA &
     RB + B4[P4+1] <= LB ] /
  P4 := P4 +1;
  RA = RA + A4[P4];
  RB = RB + B4[P4];
 }

------------------------- train 5 -----------------------------
s1 -> s1 
 {- [P5 < 6 &
     T5[P5+1] != T0[P0] &
     T5[P5+1] != T1[P1] &
     T5[P5+1] != T2[P2] &
     T5[P5+1] != T3[P3] &
     T5[P5+1] != T4[P4] &
     T5[P5+1] != T6[P6] &
     T5[P5+1] != T7[P7] &
     RA + A5[P5+1] <= LA &
     RB + B5[P5+1] <= LB] /
  P5 := P5 +1;
  RA = RA + A5[P5];
  RB = RB + B5[P5];
 }

------------------------- train 6 -----------------------------
s1 -> s1 
 {- [P6 < 6 &
     T6[P6+1] != T0[P0] &
     T6[P6+1] != T1[P1] &
     T6[P6+1] != T2[P2] &
     T6[P6+1] != T3[P3] &
     T6[P6+1] != T4[P4] &
     T6[P6+1] != T5[P5] &
     T6[P6+1] != T7[P7] &
     RA + A6[P6+1] <= LA &
     RB + B6[P6+1] <= LB ] /
  P6 := P6 +1;
  RA = RA + A6[P6];
  RB = RB + B6[P6];
 } 

------------------------- train 7 -----------------------------
s1 -> s1 
 {- [P7 < 6 &
     T7[P7+1] != T0[P0] &
     T7[P7+1] != T1[P1] &
     T7[P7+1] != T2[P2] &
     T7[P7+1] != T3[P3] &
     T7[P7+1] != T4[P4] &
     T7[P7+1] != T5[P5] &
     T7[P7+1] != T6[P6] &
     RA + A7[P7+1] <= LA &
     RB + B7[P7+1] <= LB ] /
  P7 := P7 +1;
  RA = RA + A7[P7];
  RB = RB + B7[P7];
 }

------------------------- termination -----------------------------
 s1 -> s1 
 {- [(P0=6) and (P1=6) and (P2=6) and (P3=6)&
     (P4=6) and (P5=6) and (P6=6) and (P7=6)] / ARRIVED}
 
end REGION2;

Objects:
 Count: Token;
 SYS: REGION2;

Abstractions {
 Action ARRIVED -> ARRIVED
 Action Error -> Error
-- State: 
--   SYS.P0=0 and
--   SYS.P1=0 and
--   SYS.P2=0 and
--   SYS.P3=0 and
--   SYS.P4=0 and
--   SYS.P5=0 and
--   SYS.P6=0 and
--   SYS.P7=0 -> Home  -- abstract label on final state
}

-- time umc -m3 -100 umc_oneway8.txt AFARR.txt
--
-- > The Formula: "AF {ARRIVED} true"
-- > is: TRUE 
-- > statspace stats: states generated= 1636545 ... evaluation time= 37.538 sec.
--
-- > real	0m36.980s
-- > user	1m23.800s
-- > sys	0m1.735s
-- USED VIRTUAL MEMORY: 2.98G
--
-- time mcstats -m3 umc_oneway8.txt
--
--  AFARR==   "AF {ARRIVED} true"
-------------------------------------------------------------------
-------------------------------------------------------------------


\end{verbatim}
}

\section{UPPAAL}

{\scriptsize
\begin{verbatim}
//
//  global declarations
//

//------- train missions ------
const int T0[7] = { 1, 9,10,13,15,20,23};
const int T1[7] = { 3, 9,10,13,15,20,24};
const int T2[7] = { 5,27,11,13,16,20,25};
const int T3[7] = { 7,27,11,13,16,20,26};
const int T4[7] = {23,22,17,18,11, 9, 2};
const int T5[7] = {24,22,17,18,11, 9, 4};
const int T6[7] = {25,22,17,18,12,27, 6};
const int T7[7] = {26,22,17,18,12,27, 8};

const int LA =7;  // limit value for region RA
const int LB =7;  // limit value for region RB

//------- region A: train constraints ------
const int A0[7] = { 0, 0, 0, 1, 0,-1, 0}; //G1
const int A1[7] = { 0, 0, 0, 1, 0,-1, 0}; // R1
const int A2[7] = { 0, 0, 1,-1, 0, 1, 0}; // Y1
const int A3[7] = { 0, 0, 1,-1, 0, 0, 0}; // B1
const int A4[7] = { 0, 1, 0, 0,-1, 0, 0}; // G2
const int A5[7] = { 0, 1, 0, 0,-1, 0, 0}; // R2
const int A6[7] = { 0, 0, 0,-1, 0, 0, 0}; // Y2
const int A7[7] = { 0, 1, 0,-1, 0, 0, 0}; // B2

//------- region B: train constraints ------
const int B0[7] = { 0, 0, 0, 1, 0,-1, 0}; // G1
const int B1[7] = { 0, 0, 0, 1, 0,-1, 0}; // R1
const int B2[7] = { 0, 0, 1,-1, 0, 0, 0}; // Y1
const int B3[7] = { 0, 0, 1,-1, 0, 1, 0}; // B1
const int B4[7] = { 0, 1, 0, 0,-1, 0, 0}; // G2
const int B5[7] = { 0, 1, 0, 0,-1, 0, 0}; // R2
const int B6[7] = { 0, 1, 0,-1, 0, 0, 0}; // Y2
const int B7[7] = { 0, 0, 0,-1, 0, 0, 0}; // B2
//------------------------------------------

int P0 := 0;
int P1 := 0;
int P2 := 0;
int P3 := 0;
int P4 := 0;
int P5 := 0;
int P6 := 0;
int P7 := 0;
int RA :=1;  // initial value for region RA
int RB :=1;  // initial value for region RB

broadcast chan move0,move1,move2,move3,move4,move5,move6,move7;

//------------  template defintions ---------
process Uppaal_Model() {
state s0;
urgent s0;
init s0;
trans 
  s0 -> s0 {
  guard 
    P0  < 6 &&
    T0[P0+1] != T1[P1] &&
    T0[P0+1] != T2[P2] &&
    T0[P0+1] != T3[P3] &&
    T0[P0+1] != T4[P4] &&
    T0[P0+1] != T5[P5] &&
    T0[P0+1] != T6[P6] &&
    T0[P0+1] != T7[P7] &&
    RA + A0[P0+1] <= LA  &&
    RB + B0[P0+1] <= LB;
  sync move0!;
  assign
	P0 := P0+1, 
    RA := RA + A0[P0],
    RB := RB + B0[P0];
  },
   
  s0 -> s0 {
  guard 
    P1  < 6 &&
    T1[P1+1] != T0[P0] &&
    T1[P1+1] != T2[P2] &&
    T1[P1+1] != T3[P3] &&
    T1[P1+1] != T4[P4] &&
    T1[P1+1] != T5[P5] &&
    T1[P1+1] != T6[P6] &&
    T1[P1+1] != T7[P7] &&
    RA + A1[P1+1] <= LA &&
    RB + B1[P1+1] <= LB;
  sync move1!;
  assign
	P1 := P1+1, 
    RA := RA + A1[P1],
    RB := RB + B1[P1];
  },
   
  s0 -> s0 {
  guard 
    P2 < 6 &&
    T2[P2+1] != T0[P0] &&
    T2[P2+1] != T1[P1] &&
    T2[P2+1] != T3[P3] &&
    T2[P2+1] != T4[P4] &&
    T2[P2+1] != T5[P5] &&
    T2[P2+1] != T6[P6] &&
    T2[P2+1] != T7[P7] &&
    RA + A2[P2+1] <= LA &&
    RB + B2[P2+1] <= LB;
  sync move2!;
  assign
    P2 := P2+1, 
    RA := RA + A2[P2],
    RB := RB + B2[P2];
  },
   
  s0 -> s0 {
  guard 
    P3 < 6 &&
    T3[P3+1] != T0[P0] &&
    T3[P3+1] != T2[P2] &&
    T3[P3+1] != T1[P1] &&
    T3[P3+1] != T4[P4] &&
    T3[P3+1] != T5[P5] &&
    T3[P3+1] != T6[P6] &&
    T3[P3+1] != T7[P7] &&
    RA + A3[P3+1] <= LA &&
    RB + B3[P3+1] <= LB;
  sync move3!;
  assign
    P3 := P3+1, 
    RA := RA + A3[P3],
    RB := RB + B3[P3];
  },
   
  s0 -> s0 {
  guard 
    P4 < 6 &&
    T4[P4+1] != T0[P0] &&
    T4[P4+1] != T1[P1] &&
    T4[P4+1] != T2[P2] &&
    T4[P4+1] != T3[P3] &&
    T4[P4+1] != T5[P5] &&
    T4[P4+1] != T6[P6] &&
    T4[P4+1] != T7[P7] &&
    RA + A4[P4+1] <= LA &&
    RB + B4[P4+1] <= LB;
  sync move4!;
  assign
    P4 := P4+1, 
    RA := RA + A4[P4],
    RB := RB + B4[P4];
  },

  s0 -> s0 {
  guard 
    P5 < 6 &&
    T5[P5+1] != T0[P0] &&
    T5[P5+1] != T1[P1] &&
    T5[P5+1] != T2[P2] &&
    T5[P5+1] != T3[P3] &&
    T5[P5+1] != T4[P4] &&
    T5[P5+1] != T6[P6] &&
    T5[P5+1] != T7[P7] &&
    RA + A5[P5+1] <= LA &&
    RB + B5[P5+1] <= LB;
  sync move5!;
  assign
    P5 := P5+1, 
    RA := RA + A5[P5],
    RB := RB + B5[P5];
  },
  
  s0 -> s0 {
  guard 
    P6 < 6 &&
    T6[P6+1] != T0[P0] &&
    T6[P6+1] != T1[P1] &&
    T6[P6+1] != T2[P2] &&
    T6[P6+1] != T3[P3] &&
    T6[P6+1] != T4[P4] &&
    T6[P6+1] != T5[P5] &&
    T6[P6+1] != T7[P7] &&
    RA + A6[P6+1] <= LA &&
    RB + B6[P6+1] <= LB;
  sync move6!;
  assign
    P6 := P6+1, 
    RA := RA + A6[P6],
    RB := RB + B6[P6];
  },

  s0 -> s0 {
  guard 
    P7 < 6 &&
    T7[P7+1] != T0[P0] &&
    T7[P7+1] != T1[P1] &&
    T7[P7+1] != T2[P2] &&
    T7[P7+1] != T3[P3] &&
    T7[P7+1] != T4[P4] &&
    T7[P7+1] != T5[P5] &&
    T7[P7+1] != T6[P6] &&
    RA + A7[P7+1] <= LA &&
    RB + B7[P7+1] <= LB;
  sync move7!;
  assign
    P7 := P7+1, 
    RA := RA + A7[P7],
    RB := RB + B7[P7];
  };
}

// template instantiations
// List one or more processes to be composed into a system.
system Uppaal_Model;

// ./verifyta -h
// ./verifyta uppaal_oneway8seq.ta uppaal_queries.txt
//
// file:  uppaal_queries.txt
// A<>((P0==6) and (P1==6)and (P2==6)and (P3==6) and
//     (P4==6) and (P5==6)and (P6==6)and (P7==6))
\end{verbatim}
}
\end{document}